\documentclass[sigconf]{acmart}

\usepackage{algorithm}
\usepackage{algorithmic}
\usepackage{subcaption}
\usepackage{graphicx}
\usepackage{booktabs}
\begin{document}

\title{BYOS: Knowledge-driven Large Language Models Bring Your Own Operating System More Excellent}

\author{Hongyu Lin}
\authornote{Both authors contributed equally to this research.}
\email{hongyu2021@iscas.ac.cn}
\affiliation{
\institution{University of Chinese Academy of Sciences}
\institution{Institute of Software, Chinese Academy of Sciences}
\state{Beijing}
\country{China}
}

\author{Yuchen Li}
\authornotemark[1]
\email{liyuchen2021@iscas.ac.cn}
\affiliation{
\institution{University of Chinese Academy of Sciences}
\institution{Institute of Software, Chinese Academy of Sciences}
\state{Beijing}
\country{China}
}

\author{Haoran Luo}
\authornote{Corresponding author(s).}
\email{haoran.luo@ieee.org}
\affiliation{
\institution{Nanyang Technological University}
\country{Singapore}
}

\author{Kaichun Yao}
\affiliation{
\institution{Institute of Software, Chinese Academy of Sciences}
\state{Beijing}
\country{China}
}

\author{Libo Zhang}
\affiliation{
\institution{Institute of Software, Chinese Academy of Sciences}
\state{Beijing}
\country{China}
}

\author{Zhenghong Lin}
\email{hongzhenglin970323@gmail.com}
\affiliation{
\institution{Nanyang Technological University}
\country{Singapore}
}

\author{Mingjie Xing}
\authornotemark[2]
\email{mingjie@iscas.ac.cn}
\affiliation{
\institution{University of Chinese Academy of Sciences}
\institution{Institute of Software, Chinese Academy of Sciences}
\state{Beijing}
\country{China}
}

\author{Yanjun Wu}
\affiliation{
\institution{University of Chinese Academy of Sciences}
\institution{Institute of Software, Chinese Academy of Sciences}
\state{Beijing}
\country{China}
}

\author{Carl Yang}
\email{j.carlyang@emory.edu}
\affiliation{
\institution{Emory University}
\city{Atlanta}
\state{Georgia}
\country{USA}
}

\renewcommand{\shortauthors}{Trovato et al.}

\begin{abstract}

Operating system (OS) kernel tuning is a critical yet challenging problem for performance optimization, due to the large configuration space, complex interdependencies among configuration options, and the rapid evolution of kernel versions. Recent work has explored large language models (LLMs) for automated kernel tuning, but existing approaches often suffer from hallucinated configurations, limited interpretability, and poor robustness across workloads and kernel versions. We propose BYOS, a knowledge-driven framework that grounds LLM-based Linux kernel tuning in structured domain knowledge. BYOS incorporates three key components: (1) structured knowledge construction and mapping to bridge the semantic gap, (2) knowledge-driven configuration generation to refine the search space, and (3) continuous knowledge maintenance to adapt to kernel evolution. We evaluate BYOS on diverse workloads across multiple Linux distributions and kernel versions. Experimental results show that BYOS consistently outperforms state-of-the-art tuning baselines, achieving 7.1\%–155.4\% performance improvement while substantially reducing invalid configurations. These results demonstrate the effectiveness of integrating structured knowledge with LLMs for robust and scalable system optimization. The code of BYOS is available at \url{https://github.com/LHY-24/BYOS}.
\end{abstract}

\ccsdesc[500]{Software and its engineering~Operating systems}
\ccsdesc[500]{Computing methodologies~Knowledge representation and reasoning}
\ccsdesc[300]{Information systems~Information retrieval}

\keywords{Operating System Kernel Tuning, Large Language Models, Knowledge Graph, System Performance Optimization}


\maketitle

\section{Introduction}

\begin{figure}[h]
\centering
    \includegraphics[width=0.8\linewidth]{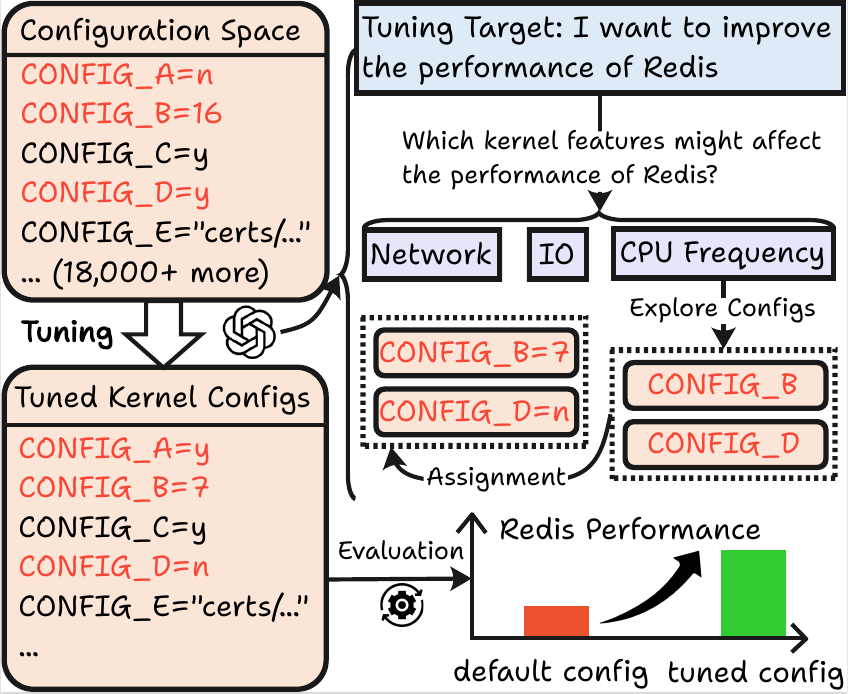}
    \caption{\fontsize{9pt}{\baselineskip}An example of kernel tuning task.}
    \Description{A conceptual diagram illustrating kernel configuration tuning for Redis performance optimization. On the left, a Linux system icon is shown with a question asking how to improve Redis performance. An arrow labeled kernel configuration options points to a kernel space box on the right that lists many configuration options such as CONFIG_A, CONFIG_B, CONFIG_C, CONFIG_D, CONFIG_E, and an indication of more than 18,000 additional options. From this kernel space, a downward arrow labeled tuning leads to a specific kernel configuration box that assigns concrete values to selected options, for example CONFIG_A set to yes, CONFIG_B set to 17, and CONFIG_D set to no. An arrow labeled eval points from the tuned configuration to a bar chart on the left. The bar chart compares system performance between a default configuration, shown as a shorter red bar, and a tuned configuration, shown as a taller green bar, indicating improved system performance after tuning.}
    \label{introduction}
\end{figure}

Operating systems (OS) serve as the critical bridge between hardware and software, forming the foundation of modern computing systems. At the core of an OS, the Linux kernel manages fundamental hardware resources, including CPU, memory, and I/O, for all running applications. Improving OS performance largely depends on effective \textbf{kernel tuning} \cite{NetOptimize1, MLOS, martin2021transfer}, which systematically adjusts kernel configuration options to optimize performance for specific workloads, as illustrated in Figure~\ref{introduction}. \par

Despite its importance, kernel tuning remains a challenging task due to the vast configuration space in modern Linux kernels, with over 18,000 configurable options and complex dependency constraints \cite{Wayfinder}, \cite{dependency}. Traditional manual tuning methods \cite{paradigm} rely heavily on expert experience, which are time-consuming and labor-intensive. While machine learning (ML)-based approaches \cite{acher, DeepPerf} offer partial automation, they often require large datasets and struggle to generalize across different hardware platforms, workload scenarios, and kernel versions.


\begin{figure*}
    \includegraphics[width=\textwidth]{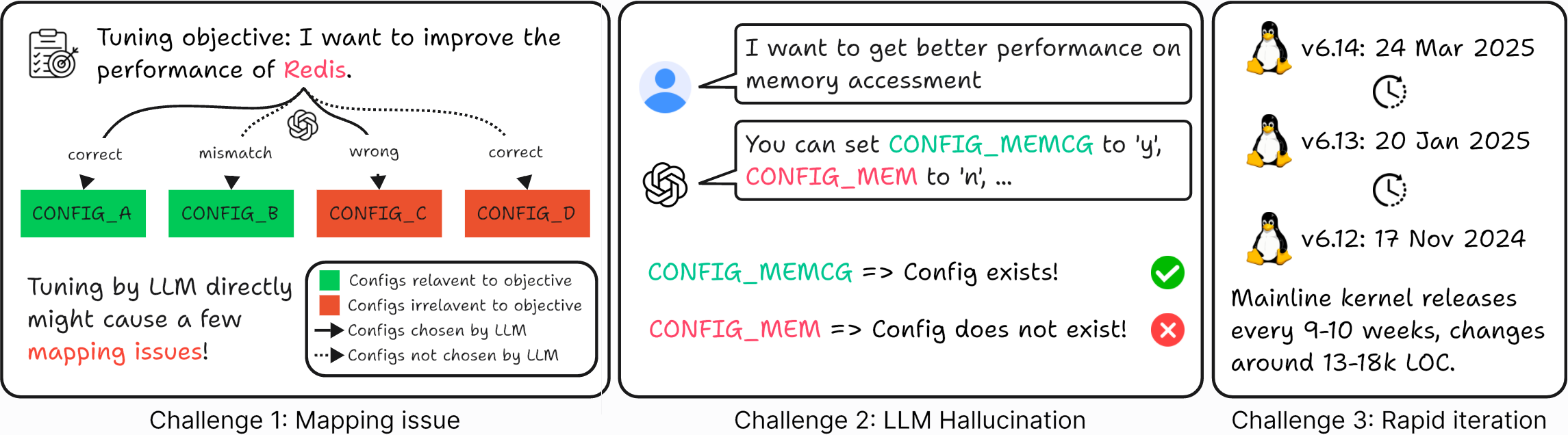}
    \caption{Challenges in OS kernel tuning with LLMs. First, LLMs struggle to map abstract tuning objectives to specific configuration options. Second, LLMs may hallucinate invalid or non-existent configurations. Third, the rapid iteration of kernel configurations, which change every few months, complicates tuning efforts.}
    \label{challenge}
\end{figure*}

Recent advances in large language models (LLMs~\cite{openai2024gpt4technicalreport,deepseekai2025deepseekv3technicalreport}) have shown promising potential for automating kernel tuning \cite{autoos}, leveraging their extensive pre-trained knowledge and strong reasoning capabilities. However, as illustrated in Figure \ref{challenge}, directly applying LLMs to kernel tuning still faces three fundamental challenges: \textbf{(1) Difficulty of mapping abstract tuning objectives to specific configuration options}: LLMs often struggle to align high-level tuning objectives expressed in natural language with the specific low-level options required for effective tuning, leading to irrelevant or suboptimal configurations. \textbf{(2) Insufficient interaction with the configuration space induces hallucinations}: without structured guidance over the vast and intricate kernel configuration space, LLMs may generate invalid, non-existent, or incompatible options \cite{reasoning}. \textbf{(3) Rapid kernel iteration causes temporal knowledge decay}: the Linux kernel evolves at a fast pace, with approximately 13k–18k commits per release and new major versions every 2–3 months \cite{Kernel}, far exceeding the ability of static LLM knowledge to remain up to date. \par

To address these challenges, we propose \textbf{BYOS} (\textbf{B}ring \textbf{Y}our own \textbf{O}perating \textbf{S}ystem more excellent), a novel knowledge-driven framework for automated Linux kernel tuning. Specifically, BYOS introduces three key innovations: \textbf{(1) Structured Knowledge Construction and Mapping.} We construct an OS-oriented dual-layer knowledge graph (OD-KG) that explicitly associate the high-level tuning objectives with corresponding low-level configuration options, enabling principled grounding of LLM reasoning. \textbf{(2) Knowledge-driven Configuration Generation.} Leveraging OD-KG, BYOS performs systematic graph-guided reasoning to generate kernel configuration, effectively constraining the search space and mitigating hallucinated or invalid outputs from LLMs. \textbf{(3) Continuous Knowledge Maintenance.} We design an efficient incremental knowledge update mechanism for OD-KG, which adapts to kernel evolution by selectively expanding and refining knowledge, thereby avoiding the need for end-to-end retraining.

We evaluate the effectiveness of BYOS using two representative OS benchmarking suites: \textit{UnixBench} \cite{unixbench} and \textit{LEBench} \cite{LEBench}, along with four widely used real-world applications: \textit{Nginx}, \textit{Redis}, \textit{Apache}, and \textit{PostgreSQL}. These applications cover diverse CPU-, memory-, storage-, and I/O-intensive workloads. Experimental results show that BYOS achieves \textbf{7.1\%-155.4\%} performance improvements over baseline methods on synthetic benchmarks and up to \textbf{42.7\%} improvement on real-world applications. These results demonstrate that BYOS provides an effective, efficient, and scalable solution for automated kernel tuning, highlighting its strong practical applicability in real-world deployment scenarios.

\section{Related Work}
\paragraph{Kernel Optimization.} Prior work on kernel optimization covers a broad range of topics. Network-specific tuning is studied in~\cite{NetOptimize1,NetOptimize2}, while \cite{MLOS} jointly optimizes kernel and application performance via machine learning. Transfer learning is leveraged to reduce kernel size in~\cite{martin2021transfer}. LEBench~\cite{LEBench} pinpoints performance regressions to configuration changes, and DeepPerf~\cite{DeepPerf} predicts performance using sparse neural networks. Kernel debloating and configuration conflicts are addressed in~\cite{Debloating} and~\cite{ConfigFix}, respectively. AutoOS~\cite{autoos} integrates LLMs with a state-machine for AIoT-oriented tuning.

\paragraph{Knowledge-Driven LLMs.} LLMs have been widely applied to software engineering tasks. Prior work studies fine-tuning for code generation~\cite{CodeSFT,CodePEFT,CodePEFT2} and vulnerability discovery~\cite{LLM_cve}. \cite{science} proposes an LLM-based code synthesis system requiring deep reasoning, while~\cite{LLM_cloud} evaluates LLMs for incident mitigation. Prompt-based configuration validation is explored in~\cite{Ciri}, and~\cite{Agentless} develops an agentless pipeline for automated bug repair.

\section{Preliminaries}
\label{sec:pre}

\paragraph{Definition 1: Configuration Space.} We model the kernel configuration space as a directed graph $\mathcal{S} = (O, E, C)$, where $O$ denotes the set of configurable options. Each option $o \in O$ is associated with an admissible value $x$ drawn from domain $D_o$. The edge set $E \subseteq O \times O$ encodes dependencies relations among options: an edge $(o_i, o_j) \in E$ indicates that $o_j$ depends on $o_i$ and cannot be configured independently. The constraint function $C : D_{o_i} \times D_{o_j} \rightarrow \{\texttt{True}, \texttt{False}\}$ specifies whether a pair of value assignments satisfies the kernel's semantic and structural constraints.


\paragraph{Definition 2: Kernel Configuration.} A kernel configuration is defined as a set $K = \{(o_1, x_1), (o_2, x_2), \dots, (o_n, x_n)\} \subseteq O \times D$, where each selected option $o_i \in O$ is assigned a value $x_i \in D_{o_i}$. A configuration $K$ is \emph{valid} if (i) all assigned values lie within their respective domains, (ii) all dependency relations induced by $E$ are satisfied, and (iii) all relevant constraints defined by $C$ are evaluated to \texttt{True} for the corresponding assignments.

\paragraph{Problem Formulation: Kernel Tuning.} Given a tuning objective $q$ and a performance evaluation function $P(K, q) \rightarrow \mathbb{R}$ that quantifies how well a configuration $K$ satisfies $q$, the kernel tuning task seeks to identify a valid configuration that maximizes $P(K,q)$ while satisfying all domain, dependency, and constraint requirements. Formally, the problem is defined as:
\[
\begin{aligned}
{\text{Maximize}} \quad & P(K,q), \quad  K \subseteq O \times D \\
\text{Subject to} \quad & x_i \in D_{o_i} \quad \forall (o_i, x_i) \in K, \\
& \texttt{Dependencies}(K, E) = \texttt{True}, \\
& \texttt{Constraints}(K, C) = \texttt{True}
\end{aligned}
\]

\begin{figure*}[ht!]
\begin{center}
\centerline{\includegraphics[width=\textwidth]{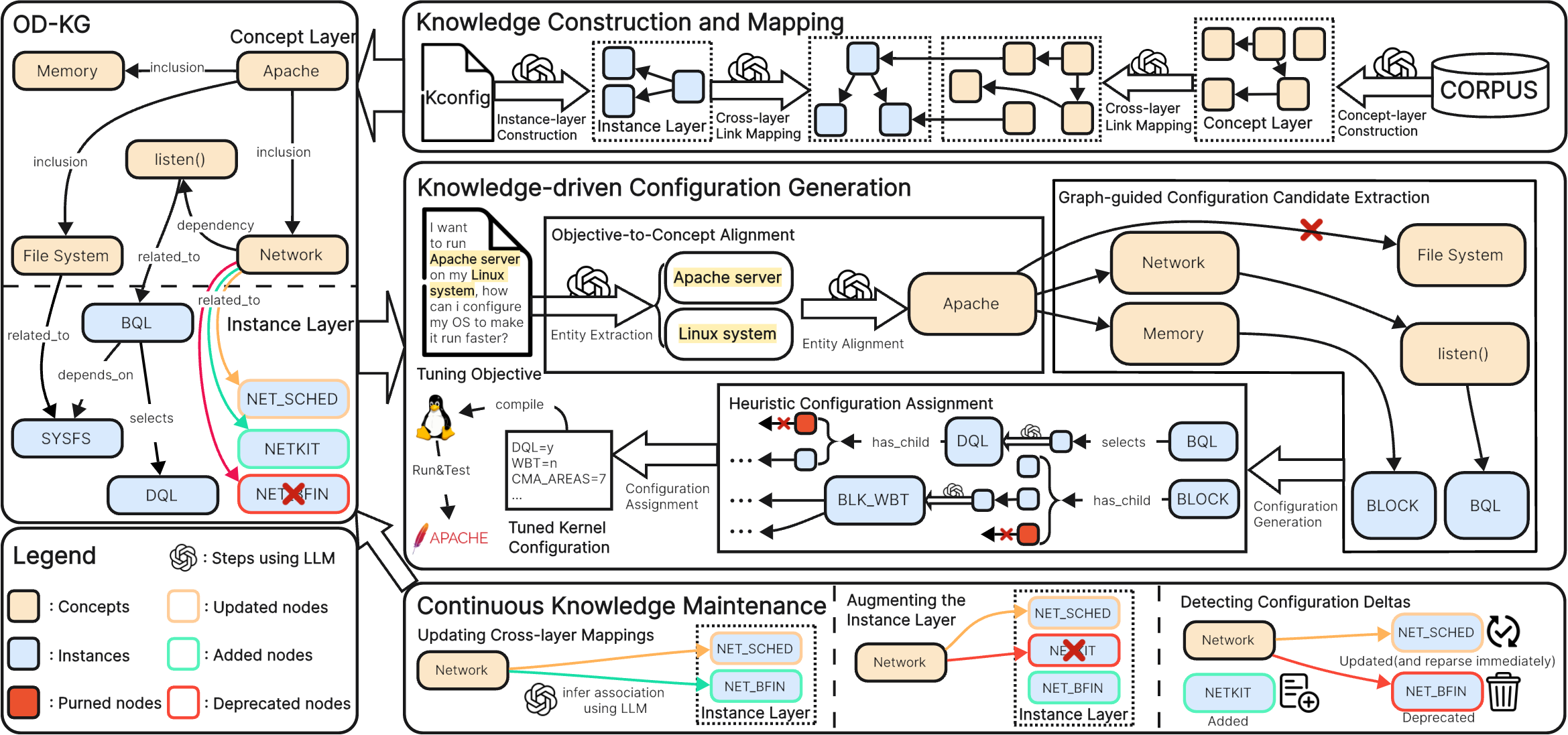}}
\caption{\fontsize{9pt}{\baselineskip}Overview of the BYOS framework for LLM-based kernel tuning. The process starts with constructing the OS-oriented Dual-layer Knowledge Graph (OD-KG), integrating tuning concepts with kernel configuration knowledge (\ref{Method_1}). Kernel configurations are then generated through knowledge-driven reasoning using the OD-KG (\ref{Method_2}). Finally, the framework supports continuous knowledge maintenance to adapt to evolving kernel configurations (\ref{Method_3}).}
\label{framework}
\end{center}
\vskip -0.2in
\end{figure*}

\section{Method: BYOS}
In this section, we introduce BYOS, a framework for LLM-based kernel tuning. As illustrated in Figure~\ref{framework}, BYOS consists of three core components: Structured Knowledge Construction and Mapping, Knowledge-driven Configuration Generation, and Continuous Knowledge Maintenance. Collectively, these components enable effective, interpretable, and robust kernel tuning.

\subsection{Knowledge Construction and Mapping}
\label{Method_1}
To address the semantic gap between high-level tuning objectives and low-level kernel configuration options, BYOS introduces a structured knowledge representation inspired by previous work on multi-level knowledge graphs \cite{JOIE,KACC,DHGE}. Specifically, BYOS constructs an \emph{OS-oriented Dual-layer Knowledge Graph (OD-KG)}, which integrates domain-level tuning concepts with kernel-specific configuration knowledge across three distinct components:
\begin{itemize}
    \item \textbf{Instance Layer.} The instance layer is defined as $\mathcal{G}_I = (\mathcal{E}_I, \mathcal{R}_I)$, where entities $\mathcal{E}_I$ represent concrete configuration options, and relations $\mathcal{R}_I$ encode dependency and structural constraints derived from the kernel configuration space.
    \item \textbf{Concept Layer.} The concept layer is defined as $\mathcal{G}_C = (\mathcal{E}_C, \mathcal{R}_C)$, where entities $\mathcal{E}_C$ correspond to abstract kernel tuning concepts, and relations $\mathcal{R}_C$ capture the semantic relationships between these concepts.
    \item \textbf{Cross-layer Links.} The cross-layer links is defined as $\mathcal{L} = \{(e_I, \allowbreak \texttt{related\_to}, \allowbreak e_C) \mid e_I \in \mathcal{E}_I, e_C \in \mathcal{E}_C\}$, which establish semantic associations between the instance-layer configuration options and concept-layer tuning objectives.
\end{itemize}
The unified OD-KG is defined as $\mathcal{G} = (V, E)$, where $V = \mathcal{E_C} \cup \mathcal{E_I}$ and $E = \mathcal{R_C} \cup \mathcal{R_I} \cup \mathcal{L}$. This dual-layer design enables interpretable reasoning from abstract tuning objectives to specific options. 

\textbf{Instance-layer Construction.} The instance layer entities $\mathcal{E}_I$ and relations $\mathcal{R}_I$ are constructed by parsing the official Linux \texttt{Kconfig} specification \cite{Kconfig}. Configuration options are extracted as entities, while dependency relations are identified through keyword-based rules. These rules cover four primary relation types defined within the Linux kernel configuration space:
\begin{equation}
\begin{aligned}
\mathcal{R}_I = \{(e_i, r, e_j) \mid\; 
& r \in \{\texttt{depends\_on}, \texttt{select}, \texttt{imply}, \texttt{has\_child}\}, \\
& e_i, e_j \in \mathcal{E}_I \}.
\end{aligned}
\end{equation}
For instance, as illustrated in Figure~\ref{fig:entity_extraction}, the option \textit{config ZSWAP} is encoded as entity $\textit{ZSWAP} \in \mathcal{E}_I$, with identified relations such as $(\textit{ZSWAP}, \texttt{depends\_on}, \textit{SWAP})$ and $(\textit{ZSWAP}, \texttt{select}, \textit{ZPOOL})$, which are captured in $\mathcal{R}_I$.

\textbf{Concept-layer Construction.} The concept layer $\mathcal{G}_C$ is constructed via few-shot in-context learning \cite{fewshot} using an LLM. Prompts (Appendix~\ref{sec:construct}) are derived from a curated corpus of kernel tuning materials, including benchmarks, research papers, and official manuals. The LLM extracts tuning objectives as entities $\mathcal{E_C}$ and infers semantic relations among them to form $\mathcal{R_C}$. Specifically, we define:
\begin{equation}
\begin{aligned}
\mathcal{R}_C = \{(e_i, r, e_j) \mid\;
& r \in \{\texttt{inclusion}, \texttt{dependency}, \texttt{influence}\}, \\
& e_i, e_j \in \mathcal{E}_C \}.
\end{aligned}
\end{equation}
For example, as shown in Figure~\ref{fig:entity_extraction}, the concepts \textit{RAM-based Memory Pool} and \textit{I/O Reduction} are linked via $(\textit{RAM-based Memory Pool}, \allowbreak \texttt{influence}, \allowbreak \textit{I/O Reduction}) \in \mathcal{R_C}$.

\textbf{Cross-layer Link Mapping.} To connect the instance layer and concept layer, BYOS leverages LLM-based semantic reasoning to establish meaningful cross-layer links. These links associate instance-layer configuration options ($\mathcal{E}_I$) with relevant tuning concepts ($\mathcal{E}_C$):  
\begin{equation}
\mathcal{L} = \{(e_I, \texttt{related\_to}, e_C) \mid e_I \in \mathcal{E}_I, e_C \in \mathcal{E}_C \}.
\end{equation}
For instance, as shown in Figure~\ref{fig:entity_extraction}, the link $(\textit{ZSWAP}, \allowbreak \texttt{related\_to}, \allowbreak \textit{Swap Pages}) \in \mathcal{L}$ illustrates the connection between a low-level option and a high-level memory tuning objective, which effectively bridges the semantic gap between them. 

\subsection{Knowledge-driven Configuration Generation}
\label{Method_2}
To reduce search overhead and mitigate LLM hallucinations, BYOS performs graph-based reasoning \cite{ToG,RoG,ChatKBQA} over the OD-KG to identify configuration options most relevant to the tuning objective, instead of exhaustive traversal-based tuning used in prior methods.

\textbf{Aligning Tuning Objectives with Kernel Concepts.} Given a tuning objective $q$, BYOS first extracts a set of textual entities $\mathcal{E}$ via semantic parsing. For example, given $q =$ \textit{``Optimize OS for faster Apache server on Linux''}, we obtain $\mathcal{E} = \{\texttt{Apache}, \texttt{Linux}\}$. Each entity $e \in \mathcal{E}$ is then mapped to a concept $e_C \in \mathcal{E}_C$ through a hybrid matching function $\phi : \mathcal{E} \rightarrow \mathcal{E}_C$. If $e \in \mathcal{E}_C$, BYOS directly identifies its corresponding concept via pattern matching $\psi_{\text{PM}}$. Otherwise, an LLM-based semantic matching function $\psi_{\text{LLM}}$ is used to rephrase $e$ and match it to the most semantically similar concept in $\mathcal{E}_C$:
\begin{equation}
\phi(e) = 
\begin{cases}
\psi_{\text{PM}}(e) & \text{if } \psi_{\text{PM}}(e) \neq \emptyset \\
\psi_{\text{LLM}}(e) & \text{otherwise}
\end{cases}
\label{eq:PM}
\end{equation}
The resulting concept set $\mathcal{E_C}^{q} = \bigcup_{e \in \mathcal{E}} \phi(e)$ captures the high-level semantics of $q$, grounding subsequent reasoning over OD-KG. \par

\textbf{Graph-guided Relevant Configuration Extraction.} For each concept $e_s \in \mathcal{E_C}^{q}$, BYOS explores reasoning paths $\pi(e_s) = \langle e_s \xrightarrow{r_1} e_1 \xrightarrow{r_2} \cdots \xrightarrow{r_n} e_n \rangle$ (where $e_i \in V$, $r_i \in E$) over OD-KG. Each path is assigned a relevance score $\rho(\pi(e_s))$, computed as:
\begin{equation}
\rho(\pi(e_s)) = \prod_{i=1}^{n} \sigma(r_i) \cdot \omega(e_i)
\end{equation}
where $\sigma(r_i)$ and $\omega(e_i)$ denote the semantic strength of relation $r_i$ and the contextual importance of node $e_i$, respectively. Configuration options reachable via paths with $\rho(\pi(e_s)) \geq \tau$ are aggregated into a candidate configuration option set $K_q$:
\begin{equation}
K_q = \left\{ e_i \in \mathcal{E_I} \,\middle|\, e_i \in \pi(e_s),\ e_s \in \mathcal{E_C}^{q},\ \rho(\pi(e_s)) \geq \tau \right\}
\end{equation}
This step effectively filters the kernel space to those options most relevant to the tuning objective.

\textbf{Heuristic Inference for Option Value Assignment.} Given the candidate set $K_q$ derived from the OD-KG, BYOS seeks to construct a complete and valid configuration $K=\{(o,x)\}\subseteq O\times D$ that maximizes the performance score $P(K,q)$ (Section~\ref{sec:pre}). To this end, BYOS iteratively assigns values to each option $o_t\in K_q$ via LLM-based inference guided by domain knowledge and structural constraints (Appendix~\ref{sec:interact}). At step $t$, the system maintains a partial configuration $K_t$ and selects an unassigned option $o_t\in K_q\setminus\{o_{t-1}\mid(o_{t-1},\cdot)\in K_t\}$. The inferred assignment is:
\begin{equation}
x_t = \texttt{LLM\_Infer}(o_t \mid \mathcal{E}_C^q, \mathcal{G}, K_t),
\end{equation}
\noindent where $\mathcal{E}_C^q$ denotes concepts aligned with $q$, $\mathcal{G}$ is the relevant OD-KG subgraph, and $K_t$ provides the current context. To ensure validity, $x_t$ must satisfy: (1) $x_t\in D_{o_t}$, (2) $\texttt{Dependencies}(K_t\cup\{(o_t,x_t)\},E)=\texttt{True}$, and (3) $\texttt{Constraints}(K_t\cup\{(o_t,x_t)\},C)=\texttt{True}$.

\textbf{Performance-aware Final Configuration Generation.} To further improve configuration quality, BYOS optionally selects the assignment $x_t^*\in D_{o_t}$ that maximizes the estimated performance score $P(K,q)$ among all valid candidates:
\begin{equation}
x_t^* = \arg\max_{x_t \in D_{o_t}} 
\left\{ P(K_t \cup \{(o_t, x_t)\}, q) \,\middle|\, 
\texttt{Valid}(K_t \cup \{(o_t, x_t)\})
\right\}.
\end{equation}
The configuration is updated as $K_{t+1}=K_t\cup\{(o_t,x_t^*)\}$, and this process repeats until all options in $K_q$ are assigned. The resulting configuration $K_T$ is both valid and semantically aligned with the tuning objective $q$ while maximizing $P(K,q)$. The full algorithmic procedure is described in Appendix~\ref{sec:alg}:
\begin{equation}
K_T = \bigcup_{t=0}^{T-1} \{(o_t,x_t^*)\}.
\end{equation}

\subsection{Continuous Knowledge Maintenance}
\label{Method_3}
To keep pace with the rapid evolution of the Linux kernel and maintain up-to-date knowledge, BYOS adopts an incremental update strategy to continuously refine the OD-KG with minimal overhead. Let $S^{(t)}=(O^{(t)},E^{(t)},C^{(t)})$ denote the kernel space at version $t$ as defined in Section~\ref{sec:pre}, and $S^{(t+1)}$ the subsequent version. Our goal is to update the instance layer $\mathcal{G}_I^{(t+1)}$ and the cross-layer links $\mathcal{L}^{(t+1)}$ to reflect $S^{(t+1)}$, while preserving the existing concept-layer $\mathcal{G}_C^{(t)}$.

\textbf{Step 1: Detecting Configuration Deltas.} We compute the configuration delta between two consecutive kernel versions to capture changes in the option set. Newly added options are defined as $\Delta O_{\text{add}}=\{o\in O^{(t+1)}\mid o\notin O^{(t)}\}$, and deprecated options as $\Delta O_{\text{del}}=\{o\in O^{(t)}\mid o\notin O^{(t+1)}\}$. For options that persist across versions but exhibit changes in their domains or dependencies, we re-parse and update their corresponding entities and relations to reflect the latest semantics.

\textbf{Step 2: Augmenting the Instance Layer.} For each new option $o \in \Delta\mathcal{O}_{\text{add}}$, we add its corresponding new entity $e_o$ to $\mathcal{E_I}^{(t+1)}$, then extract its relations $\mathcal{R}_o$ and insert them into $\mathcal{R_I}^{(t+1)}$:
\begin{equation}
\begin{aligned}
\mathcal{R}_o
= \{ (e_o, r, e_{o'}) \mid\;
& r \in \{\texttt{depends\_on}, \texttt{select},
\texttt{imply}, \texttt{has\_child}\}, \\
& e_{o'} \in \mathcal{O}^{(t+1)} \}.
\end{aligned}
\end{equation}
For each deprecated option $o \in \Delta\mathcal{O}_{\text{del}}$, we delete the associated entity $e_o$ and all related relations:
\begin{align}
\mathcal{E_I}^{(t+1)} &\leftarrow \mathcal{E_I}^{(t)} \setminus \{ e_o \mid o \in \Delta\mathcal{O}_{\text{del}}\} \\
\mathcal{R_I}^{(t+1)} &\leftarrow \mathcal{R_I}^{(t)} \setminus \{ (e_{o_i}, r, e_{o_j}) \mid e_{o_i} = e_o \lor e_{o_j} = e_o \}
\end{align}

\textbf{Step 3: Updating Cross-layer Mappings.} For each new instance entity $e_I \in \mathcal{E_I}^{(t+1)} \setminus \mathcal{E_I}^{(t)}$, we invoke a LLM to infer its semantic association with concept-layer entities $e_C \in \mathcal{E_C}$, forming new cross-layer links:
\begin{equation}
\Delta \mathcal{L}^{(t+1)} = \{ (e_I, \texttt{related\_to}, e_C) \}
\end{equation}
These links ensure that newly introduced kernel options remain interpretable through high-level domain knowledge mapping. Deprecated options have their cross-layer links removed accordingly.

\begin{table*}[htbp]
\small
\centering
\caption{Best UnixBench results across four Linux distributions (higher is better).
Bold numbers indicate the best score within each distribution, and percentages denote relative improvement over the default configuration. ET = Execl Throughput, FC = File Copy, PT = Pipe Throughput, CS = Context Switching, 
PC = Process Creation, SS = Shell Scripts, and SC = System Call.}
\label{unixbench_results}
\begin{tabular}{@{}c cccccccccccc c@{}} \toprule & Dhrystone & Whetstone & ET & FC 1024 & FC 256 & FC 4096 & PT & CS & PC & SS 1 & SS 8 & SC & Total Score \\ 
\midrule
\multicolumn{14}{c}{\textbf{Ubuntu}} \\
\midrule
Default & 5182 & 1842 & 1489 & 5466 & 3863 & 9629 & 2866 & 864 & 1145 & 4205 & 9003 & 2529 & 3099 \\
Qwen3-7B & 5495 & 1818 & 1504 & 5971 & 3564 & 9564 & 2587 & 802 & 1159 & 4069 & 8705 & 2219 & 3010 (-2.9\%) \\
DeepSeek-R1 & 5538 & 1817 & 1497 & 5937 & 3815 & 9257 & 2752 & 850 & 1167 & 4219 & 9046 & 2533 & 3120 (+0.7\%) \\
GPT-5 & 5389 & 1826 & 1530 & 5879 & 3781 & 9596 & 2843 & 862 & 1172 & 4277 & 8923 & 2373 & 3115 (+0.5\%) \\
AutoOS & 5616 & \textbf{1864} & 1533 & 5976 & 3819 & 9458 & 2945 & 854 & 1150 & 4241 & 9032 & 2527 & 3154 (+1.8\%) \\
\textbf{BYOS (Qwen3-7B)} & 5320 & 1807 & 1702 & 5925 & 3742 & 9751 & 2788 & 947 & 1347 & \textbf{4664} & \textbf{9932} & 2319 & 3242 (+4.6\%) \\
\textbf{BYOS (DeepSeek-R1)} & \textbf{5716} & 1796 & \textbf{1704} & 5741 & 3768 & 9687 & 2835 & \textbf{1003} & \textbf{1373} & 4552 & 9875 & 2485 & 3289 (+6.1\%) \\
\textbf{BYOS (GPT-5)} & 5525 & 1848 & 1628 & \textbf{6266} & \textbf{4105} & \textbf{10079} & \textbf{3091} & 897 & 1231 & 4587 & 9816 & \textbf{2684} & \textbf{3318 (+7.1\%)} \\
\midrule
\multicolumn{14}{c}{\textbf{Fedora}} \\
\midrule
Default & 4706 & 1617 & 233 & 1022 & 701 & 2290 & 416 & 51 & 332 & 991 & 3526 & 157 & 689 \\
Qwen3-7B & \textbf{5049} & \textbf{1692} & 250 & 1147 & 678 & 2352 & 399 & 52 & 338 & 1017 & 3467 & 156 & 705 (+2.3\%) \\
DeepSeek-R1 & 4759 & 1631 & 243 & 1097 & 754 & 2510 & 452 & 149 & 390 & 1143 & \textbf{4739} & 168 & 821 (+19.2\%) \\
GPT-5 & 4891 & 1675 & 258 & 1172 & 781 & 2535 & 462 & 160 & 406 & 1156 & 4473 & 175 & 845 (+22.7\%) \\
AutoOS & 4969 & 1669 & \textbf{281} & 1302 & 833 & 2613 & 458 & 147 & 397 & 1078 & 3981 & 177 & 846 (+22.8\%) \\
\textbf{BYOS (Qwen3-7B)} & 4959 & 1662 & 264 & 1162 & 790 & 2535 & 463 & 168 & 389 & 1089 & 3758 & 175 & 832 (+20.8\%) \\
\textbf{BYOS (DeepSeek-R1)} & 5021 & 1675 & 250 & 1317 & 875 & 2844 & 524 & 204 & \textbf{453} & \textbf{1179} & 4424 & 209 & 919 (+33.4\%) \\
\textbf{BYOS (GPT-5)} & 4871 & 1689 & 258 & \textbf{1319} & \textbf{922} & \textbf{2885} & \textbf{558} & \textbf{239} & 400 & 1155 & 4542 & \textbf{217} & \textbf{936 (+35.9\%)} \\
\midrule
\multicolumn{14}{c}{\textbf{Debian}} \\
\midrule
Default & 6271 & \textbf{2044} & 1315 & 5031 & 3162 & 10029 & 2300 & 276 & 1199 & 4689 & 10702 & 1604 & 2721 \\
Qwen3-7B & 6260 & 2036 & 1198 & 5099 & 3198 & 10261 & 2169 & 227 & 1091 & 4515 & 10139 & 1261 & 2560 (-5.9\%) \\
DeepSeek-R1 & 6279 & 2033 & 1316 & 5785 & 3711 & 11116 & 2686 & 402 & \textbf{1253} & 4814 & 11044 & 2007 & 3022 (+11.1\%) \\
GPT-5 & 6066 & 2002 & 1131 & 5584 & 3569 & 10473 & 2572 & 344 & 1045 & 4278 & 9671 & 1959 & 2782 (+2.2\%) \\
AutoOS & 6346 & 2041 & 1356 & 6646 & 4143 & 12070 & 2964 & 405 & 1209 & 4715 & 10695 & 2404 & 3169 (+16.5\%) \\
\textbf{BYOS (Qwen3-7B)} & 6121 & 2007 & 1184 & \textbf{7571} & \textbf{4955} & 13264 & \textbf{3598} & 384 & 1146 & 4714 & 10841 & 2307 & 3243 (+19.2\%) \\
\textbf{BYOS (DeepSeek-R1)} & \textbf{6437} & 2031 & \textbf{1392} & 7235 & 4557 & 13178 & 3417 & 510 & 1235 & \textbf{4930} & \textbf{11134} & \textbf{2421} & \textbf{3385 (+24.4\%)} \\
\textbf{BYOS (GPT-5)} & 6298 & 2035 & 1221 & 7538 & 4896 & \textbf{13828} & 3522 & \textbf{514} & 1098 & 4531 & 10385 & 2273 & 3305 (+21.5\%) \\
\midrule
\multicolumn{14}{c}{\textbf{OpenEuler}} \\
\midrule
Default & 3442 & 1300 & 210 & 614 & 372 & 1565 & 240 & 42 & 88 & 441 & 3650 & 123 & 442 \\
Qwen3-7B & 3470 & 1291 & \textbf{332} & 603 & 365 & 1530 & 227 & 41 & 74 & 380 & 2585 & 121 & 430 (-2.7\%) \\
DeepSeek-R1 & 3496 & \textbf{1543} & 241 & 501 & 464 & 1795 & 294 & \textbf{61} & 170 & 674 & 1164 & 332 & 540 (+22.2\%) \\
GPT-5 & 3497 & 1412 & 189 & 939 & 582 & 2090 & 698 & 33 & 201 & 740 & 972 & 590 & 599 (+35.5\%) \\
AutoOS & 3164 & 1200 & 237 & 2960 & 1989 & 6302 & 1393 & 40 & 107 & 603 & 3955 & 1071 & 945 (+113.8\%) \\
\textbf{BYOS (Qwen3-7B)} & \textbf{3665} & 1456 & 205 & 1007 & 633 & 2226 & 616 & 47 & \textbf{229} & \textbf{825} & 1048 & 596 & 646 (+46.2\%) \\
\textbf{BYOS (DeepSeek-R1)} & 3490 & 1302 & 237 & \textbf{3753} & \textbf{2515} & \textbf{7506} & \textbf{1662} & 47 & 119 & 632 & \textbf{4332} & 1627 & 1101 (+149.1\%) \\
\textbf{BYOS (GPT-5)} & 3500 & 1315 & 251 & 3674 & 2405 & 7323 & 1635 & 54 & 135 & 648 & 4256 & \textbf{1643} & \textbf{1129 (+155.4\%)} \\
\bottomrule
\end{tabular}
\end{table*}

\section{Experiments}
\label{sec:exp}
We evaluate the effectiveness and efficiency of BYOS through a comprehensive set of studies designed to answer the following research questions (RQs): \textbf{RQ1:} How does BYOS compare with existing state-of-the-art baselines for kernel tuning? \textbf{RQ2:} How does the key components of BYOS contribute to its overall performance? \textbf{RQ3:} How effectively does BYOS address the knowledge mapping challenge? \textbf{RQ4:} To what extent does BYOS mitigate hallucinated or invalid configurations generated by LLMs? \textbf{RQ5:} Can BYOS maintain tuning effectiveness across evolving kernel versions? \textbf{RQ6:} How does BYOS perform in real-world applications scenarios? \textbf{RQ7:} What is the tuning cost of BYOS in terms of computational and inference overhead?

\begin{table*}[htbp!]
\small
\centering
\caption{Ablation study of GPT-5-based BYOS on Ubuntu using the UnixBench benchmark. \textbf{Default} refers to the system’s default configuration; \textbf{w/o KG} removes the OD-KG knowledge base; and \textbf{w/o Mapping} removes the structured knowledge mapping strategy. Abbreviations of sub-tests are consistent with Table~\ref{unixbench_results}.}
\label{tab:ablation}
\begin{tabular}{@{}l cccccccccccc c@{}}
\toprule
Variant & Dhrystone & Whetstone & ET & FC 1024 & FC 256 & FC 4096 & PT & CS & PC & SS 1 & SS 8 & SC & Total Score \\
\midrule
Default & 5182 & 1842 & 1489 & 5466 & 3863 & 9629 & 2866 & 864 & 1145 & 4205 & 9003 & 2529 & 3099 \\
w/o Mapping & 5495 & 1818 & 1504 & 5710 & 3564 & 9564 & 2587 & 802 & 1159 & 4069 & 8705 & 2219 & 3010 (-2.9\%) \\
w/o KG & 5389 & 1826 & 1530 & 5879 & 3781 & 9596 & 2843 & 862 & 1172 & 4277 & 8923 & 2373 & 3115 (+0.5\%) \\
BYOS (GPT-5)
& \textbf{5525} & \textbf{1848} & \textbf{1628} & \textbf{6266} & \textbf{4105} & \textbf{10079}
& \textbf{3091} & \textbf{897} & \textbf{1231} & \textbf{4587} & \textbf{9816} & \textbf{2684}
& \textbf{3318 (+7.1\%)} \\
\bottomrule
\end{tabular}
\end{table*}

\subsection{Experimental Setup}
\textbf{Linux Distributions.} 
To ensure a comprehensive and representative evaluation across diverse system usage scenarios, we consider four widely adopted Linux distributions: \textbf{Ubuntu}, \textbf{Fedora}, \textbf{Debian}, and \textbf{OpenEuler}. Table~\ref{tab:OS} summarizes the detailed specifications of each distribution used in our experiments.

\begin{table}[htbp!]
\centering
\tabcolsep=1.0mm
\caption{Details of four representative Linux distributions.}
\label{tab:OS}
\begin{tabular}{@{}cccc@{}}
\toprule
\textbf{Distribution} & \textbf{Version} & \textbf{Kernel} & \textbf{Main Scenario} \\ 
\midrule
Ubuntu    & 22.04           & Linux 6.2.16  & Desktop, Server, IoT \\
Fedora    & 41              & Linux 6.2.16  & Development \& Test \\
Debian    & 12              & Linux 6.1.45  & Embedded System \\ 
OpenEuler & 22.03           & Linux 6.6.45  & Cloud Computing, AI\\
\bottomrule
\end{tabular}
\end{table}

\textbf{Benchmarks.}
We use two standard OS benchmarking suites to assess kernel performance. Specifically, \textbf{UnixBench}~\cite{unixbench} serves as a macro-benchmark that aggregates multiple sub-tests (e.g., context switching and pipe throughput) to measure overall system performance, while \textbf{LEBench}~\cite{LEBench} functions as a micro-benchmark that evaluates fine-grained critical kernel operations at system-call level (e.g., \texttt{fork} and \texttt{mmap}).

\textbf{Applications.}
To further evaluate the impact of kernel tuning on real-world workloads, we consider four representative applications: \textbf{Nginx Web Server}~\cite{Nginx}, \textbf{Apache HTTP Server}~\cite{Apache}, \textbf{Redis Key-Value Store}~\cite{Redis}, and \textbf{PostgreSQL Database}~\cite{PostgreSQL}. Nginx and Apache are evaluated using ApacheBench~\cite{apache_bench}, Redis using Redis Benchmark~\cite{redis_benchmarks}, and PostgreSQL using sysbench~\cite{sysbench}.

\textbf{Hardware.}
All experiments are conducted on a dual-socket workstation equipped with Intel Xeon Gold 6430 CPUs (64 cores, 128 threads), 1 TB of main memory, and 8 GB of swap space, running a 64-bit Linux operating system.

\textbf{Baselines.}
We compare \textsc{BYOS} against three representative baselines: 
(1) \textbf{Default Configuration}, which reflects commonly used expert-driven kernel settings in practice; 
(2) \textbf{Vanilla LLM Tuning}, which directly applies GPT-5, DeepSeek-R1, and Qwen3-7B using tuning workflows and prompt templates identical to those of BYOS; and 
(3) \textbf{AutoOS Tuning}~\cite{autoos}, a state-of-the-art LLM-based kernel tuning framework that leverages a predefined state machine to iteratively guide configuration decisions.

\textbf{Implementation.}
All methods are evaluated under identical settings. Each experiment is repeated for at least 30 independent runs per objective, and the best-performing configuration is reported, consistent with the goal of kernel tuning to identify a high-quality configuration through iterative exploration. For fairness, although AutoOS by default uses the OpenAI API with GPT-4o-mini, we replace its backbone model with GPT-5 in our experiments; moreover, we report the best performance among configurations generated by our AutoOS runs and those publicly released by AutoOS.

\subsection{Overall Kernel Performance (RQ1)}
\label{subsec:RQ1}
Table~\ref{unixbench_results} presents a comparison between BYOS and representative baselines under different backbone models. Across all evaluated Linux distributions, BYOS achieves the highest overall UnixBench scores, with improvements of up to \textbf{7.1\%} on Ubuntu, \textbf{35.9\%} on Fedora, \textbf{24.4\%} on Debian, and \textbf{155.4\%} on openEuler over the default configurations, demonstrating strong effectiveness and robustness across heterogeneous systems.

\textbf{BYOS consistently strengthens LLM-based tuning across model scales.} Compared to directly applying LLMs, BYOS yields consistent performance gains when instantiated with Qwen3-7B, GPT-5, and DeepSeek-R1. For example, on Ubuntu, vanilla LLM tuning shows marginal gains or even degradation (e.g., $-2.9\%$ for Qwen3-7B), whereas BYOS improves performance by \textbf{4.6\%–7.1\%}. Similar trends are observed across Fedora, Debian, and openEuler, indicating that the gains stem from the proposed knowledge-driven tuning workflow rather than any specific model.

\textbf{Larger models further benefit from the BYOS workflow due to stronger reasoning and knowledge utilization capabilities.} BYOS instantiations built on larger models (e.g., GPT-5 and DeepSeek-R1) generally achieve higher overall scores than those based on Qwen3-7B, suggesting that increased model capacity better exploits structured kernel knowledge and BYOS is complementary to model scaling across different parameter regimes.

\textbf{BYOS prioritizes overall system performance by effectively balancing competing sub-objectives.} Although baselines may outperform BYOS on certain isolated sub-tests (e.g., Whetstone on Ubuntu), BYOS consistently achieves stronger aggregate performance across distributions, demonstrating its ability to mitigate conflicts among kernel sub-modules and to optimize system performance holistically.

\subsection{Ablation Study (RQ2)}
\label{subsec:RQ2}
As shown in Table~\ref{tab:ablation}, we conduct an ablation study on Ubuntu using GPT-5 to quantify the contribution of individual BYOS components.

\textbf{Effect of Knowledge Mapping.} Removing the knowledge mapping module (w/o Mapping) causes a clear performance degradation (\textbf{-2.9\%} vs. default and \textbf{-6.6\%} vs. BYOS). Without explicit alignment between tuning objectives and relevant configuration options, the LLM produces less coherent and objective-aligned decisions, highlighting the importance of knowledge mapping for grounding domain-specific tuning.

\textbf{Effect of OD-KG.} Disabling the OD-KG (w/o KG) yields only marginal improvement over the default configuration (\textbf{+0.5\%}) and remains far inferior to full BYOS (\textbf{+7.1\%}). This suggests that while the LLM can infer coarse-grained intent, structured domain knowledge is essential for reasoning about configuration dependencies and identifying high-impact optimizations.

\textbf{Overall contribution.} The full BYOS system achieves the highest UnixBench score (\textbf{3318}, \textbf{+7.1\%}), surpassing both the default configuration and all ablated variants. This result suggests that the gains arise from the synergistic effect of structured mapping and knowledge-driven reasoning, rather than from the LLM alone.

\begin{figure*}[htbp]
    \begin{center}
    \includegraphics[width=\textwidth]{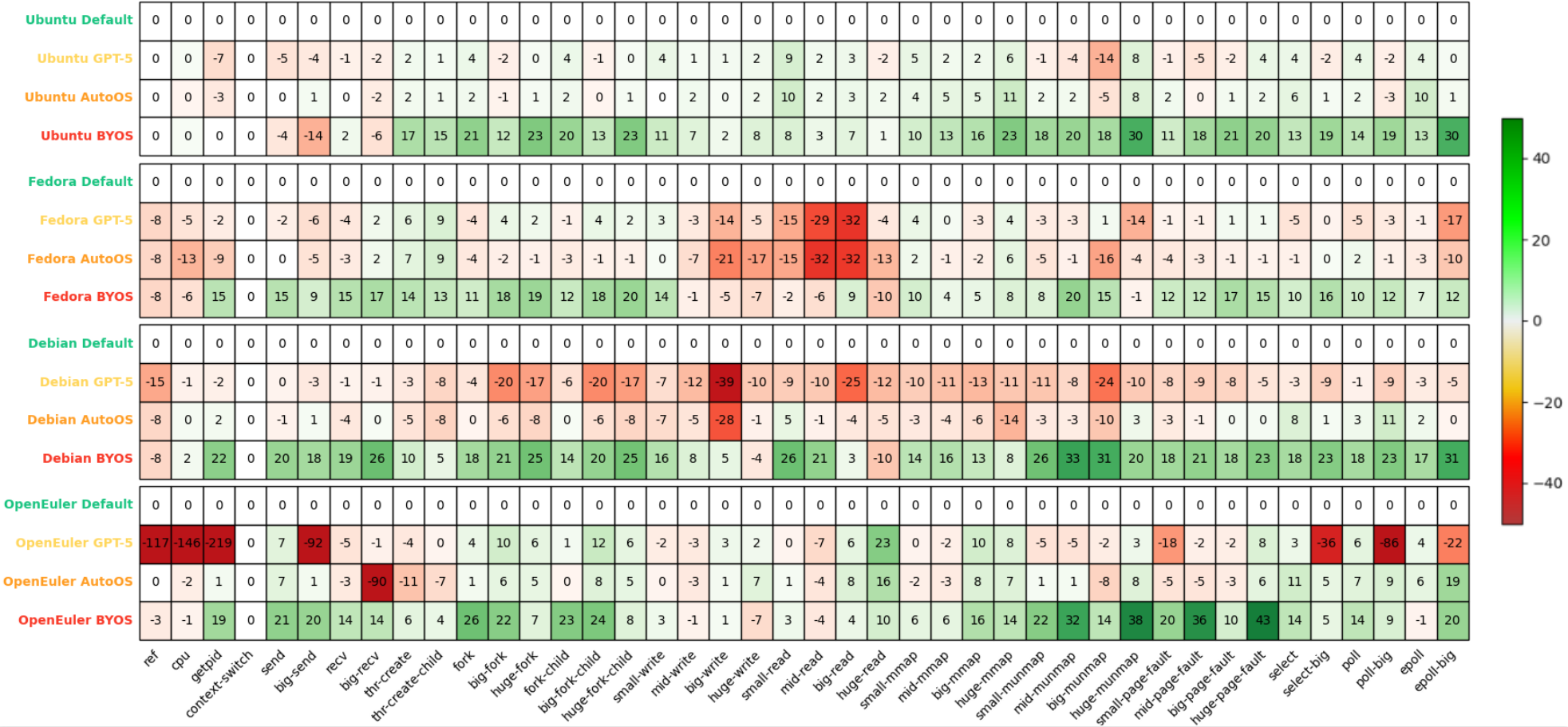}
    \caption{Result of LEBench: the heatmap shows the relative latency changes in kernel operations for each method, compared to the default configuration. Green indicates reduced latency (better), while red denotes increased latency (worse).}
    \label{lebench-result}
    \end{center}
\end{figure*}


\begin{table*}[htbp!]
\caption{Robustness comparison on tuning Ubuntu over 10 independent runs using UnixBench scores. We report per-run results, success rate, failure breakdown (compile error (CE) / boot error (BE)), variance ($\sigma^2$) over valid runs, and the best score achieved. BYOS uses GPT-5, and higher scores indicate better performance.}
\centering
\begin{tabular}{lcccccccccc|cc|cc}
\toprule
 & obj1 & obj2 & obj3 & obj4 & obj5 & obj6 & obj7 & obj8 & obj9 & obj10
 & Success Rate & CE/BE & Var ($\sigma^2$) & Best \\
\midrule
AutoOS 
& 2976 & 1779 & CE & BE & 3154 & CE & 3087 & CE & BE & 2660
& 50.0\% & 3/2 & 255{,}408 & 3154 \\
BYOS (GPT-5) 
& 3025 & 3113 & 2883 & BE & 3261 & 3164 & 2924 & 3307 & CE & 3318
& 80.0\% & 1/1 & 24{,}939 & 3318 \\
\bottomrule
\end{tabular}
\label{stability}
\end{table*}

\subsection{Structured Knowledge Mapping Enhances Fine-Grained Tuning (RQ3)}
\label{subsec:RQ3}
As shown in Figure~\ref{lebench-result}, we evaluate BYOS's ability to map high-level tuning objectives to low-level configuration options using LEBench, with GPT-5 as the underlying LLM, which measures system-call latency and thus captures fine-grained kernel behavior. \par

\textbf{BYOS consistently improves fine-grained kernel operations.} BYOS reduces latency across most evaluated system calls, including \texttt{fork}, \texttt{thr-create}, \texttt{mmap}, \texttt{page-fault}, and \texttt{epoll}, indicating effective identification and optimization of tuning-relevant configuration options rather than indiscriminate tuning. \par

\textbf{Baseline methods exhibit inconsistent optimization behavior.} In contrast, both AutoOS and vanilla LLM-based approaches show mixed effects, improving some system calls while degrading others. This instability reflects the lack of structured guidance for determining which options are relevant to a given fine-grained objective, leading to suboptimal trade-offs among kernel operations. \par

Overall, these results demonstrate that \textbf{structured knowledge mapping in BYOS enables precise and objective-aligned tuning at the system call level}, effectively bridging the semantic gap between high-level objective and low-level configurations. \par

\subsection{Mitigating LLM Hallucinations (RQ4)}
\label{subsec:RQ4}
Table~\ref{stability} evaluates robustness to LLM hallucinations via 10 independent tuning runs on Ubuntu. Each run is labeled as \textbf{CE} (compile error), \textbf{BE} (boot error), or a valid \textbf{Score} if the kernel successfully boots and completes UnixBench.

\textbf{BYOS significantly improves configuration validity and stability.}
BYOS achieves a higher success rate (\textbf{80.0\%}), with only one compile error and one boot failure across all runs. Among valid runs, BYOS exhibits substantially lower performance variance (\textbf{24,939}) and attains a higher best UnixBench score (\textbf{3318}). These results indicate that BYOS’s knowledge-driven reasoning effectively mitigates LLM hallucinations and enhances configuration validity.

\textbf{AutoOS exhibits frequent invalid configurations and unstable performance.}
AutoOS succeeds in only 50.0\% of runs, with three compile errors and two boot failures. Even among valid runs, performance fluctuates considerably, yielding a high variance (\textbf{255,408}), which reflects unstable tuning behavior.

Overall, these findings demonstrate that \textbf{the knowledge-driven configuration generation in BYOS substantially mitigates LLM hallucinations in kernel tuning}, leading to more valid and stable configurations.

\begin{figure*}[htbp]
    \includegraphics[width=\textwidth]{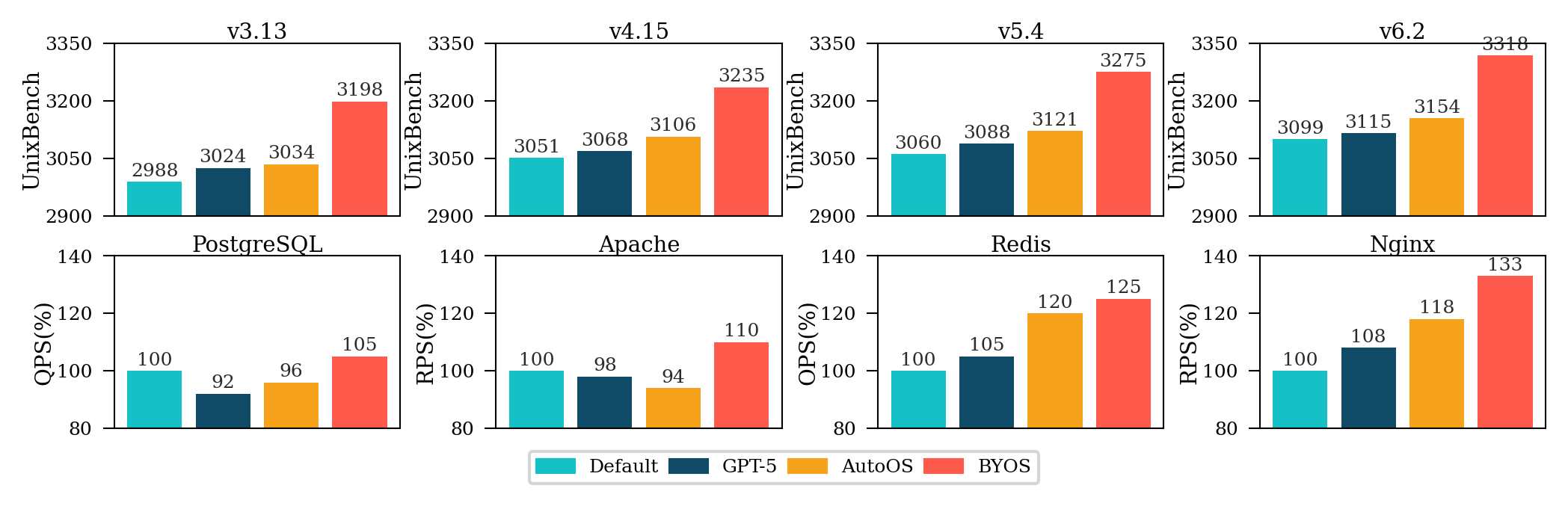}
    \caption{Performance comparison across Linux kernel versions (top row) and real-world applications (bottom row). For each kernel version (v3.13, v4.15, v5.4, and v6.2), we report the absolute UnixBench score under four configurations: Default, GPT-5, AutoOS, and BYOS. For real-world workloads, performance is normalized to the default configuration, where Nginx and Apache are measured in \textbf{Requests per Second (RPS)}, Redis in \textbf{Operations per Second (OPS)}, and PostgreSQL in \textbf{Queries per Second (QPS)}.}
    \label{app-result}
\end{figure*}

\subsection{Adaptability Across Kernel Versions (RQ5)}
\label{subsec:RQ5}
As shown in Figure~\ref{app-result}, we assess the adaptability of BYOS across Linux kernel versions by applying it to four Ubuntu releases (14.04, 16.04, 18.04, and 20.04), corresponding to kernel versions 3.13, 4.15, 5.4, and 6.2, with performance evaluated using UnixBench.

\textbf{BYOS consistently improves performance across all evaluated kernel versions}, achieving gains ranging from \textbf{3.7\%} to \textbf{12.1\%} despite substantial changes in the configuration space. This robustness stems from the continuous knowledge maintenance mechanism in OD-KG, which tracks upstream kernel evolution and incrementally updates tuning-relevant knowledge.

Overall, these results indicate that \textbf{continuous knowledge maintenance in BYOS effectively adapts to kernel evolution}, making it a sustainable solution for long-term kernel tuning.

\subsection{Real-world Application Evaluation (RQ6)}
\label{subsec:RQ6}
As shown in Figure~\ref{app-result}, we evaluate the practical effectiveness of BYOS under real-world conditions using four widely deployed applications—Nginx, Apache, Redis, and PostgreSQL—spanning web serving, database, and in-memory data processing workloads with diverse CPU, I/O, and memory characteristics.

\textbf{BYOS consistently outperforms both the default configuration and AutoOS across all applications}. In particular, Redis throughput improves by up to \textbf{25.0\%}, while Nginx achieves latency reductions of up to \textbf{42.7\%}. These improvements are attributed to BYOS’s rich application-aware knowledge, which allows it to identify kernel optimization patterns most relevant to each workload.

Overall, the results demonstrate that \textbf{BYOS effectively translates kernel-level optimizations into tangible application-level performance gains}, validating its effectiveness and generality in real-world deployment scenarios.

\begin{table}[htbp]
\centering
\caption{Tuning Cost Statistics (\textit{mean} $\pm$ \textit{std})}
\label{tab:cost}
\setlength{\tabcolsep}{7pt}
\renewcommand{\arraystretch}{1.2}
\begin{tabular}{l rr}
\toprule
\textbf{Metric} & \textbf{AutoOS} & \textbf{BYOS} \\
\midrule
\quad Runtime (s)          
& $1122.4 \pm 265.0$ 
& $\mathbf{594.3 \pm 96.1}$  \\

\quad API Calls      
& $145.8 \pm 30.4$  
& $244.0 \pm 20.5$  \\

\quad Prompt Tokens        
& $119{,}328 \pm 26{,}405$ 
& $468{,}198 \pm 63{,}208$ \\

\quad Completion Tokens    
& $47{,}523 \pm 10{,}40$  
& $\mathbf{15{,}694 \pm 2{,}106}$   \\
\bottomrule
\end{tabular}
\end{table}

\subsection{Tuning Cost Analysis (RQ7)}
\label{subsec:RQ7}
As shown in Table~\ref{tab:cost}, to analyze tuning cost and efficiency, we record the average runtime, number of inference calls, and token usage of BYOS and AutoOS under identical objectives with multiple times independent executions.

\textbf{Knowledge-driven reasoning enables faster convergence.}
BYOS reduces tuning time by nearly \textbf{47\%} compared to AutoOS (594.36s vs. 1122.39s) by reasoning over OD-KG to precisely identify tuning-relevant kernel configurations and substantially reduce the effective search space. In contrast, AutoOS relies on traversal-based exploration, leading to lower tuning efficiency.

\textbf{Richer prompts yield more focused generations.}
Although BYOS consumes more prompt tokens due to knowledge-grounded context, its completion token usage is reduced by \textbf{3$\times$} (15.7k vs. 47.5k), indicating BYOS achieves a more effective tuning process by trading richer semantic grounding for better performance.

\section{Conclusion}
We present \textbf{BYOS}, a knowledge-driven framework that leverages LLMs for OS kernel tuning. By integrating a dual-layer, OS-centric knowledge graph with targeted retrieval, BYOS effectively bridges high-level tuning objectives and low-level kernel configuration options. Extensive experiments demonstrate that BYOS consistently outperforms existing approaches in terms of performance, stability, and adaptability across kernel versions and real-world workloads. Overall, this work highlights the broader potential of structured knowledge integration for reliable and efficient LLM-based system software optimization.

\bibliographystyle{ACM-Reference-Format}
\bibliography{sample-base}


\begin{thebibliography}{42}


\ifx \showCODEN    \undefined \def \showCODEN     #1{\unskip}     \fi
\ifx \showISBNx    \undefined \def \showISBNx     #1{\unskip}     \fi
\ifx \showISBNxiii \undefined \def \showISBNxiii  #1{\unskip}     \fi
\ifx \showISSN     \undefined \def \showISSN      #1{\unskip}     \fi
\ifx \showLCCN     \undefined \def \showLCCN      #1{\unskip}     \fi
\ifx \shownote     \undefined \def \shownote      #1{#1}          \fi
\ifx \showarticletitle \undefined \def \showarticletitle #1{#1}   \fi
\ifx \showURL      \undefined \def \showURL       {\relax}        \fi
\providecommand\bibfield[2]{#2}
\providecommand\bibinfo[2]{#2}
\providecommand\natexlab[1]{#1}
\providecommand\showeprint[2][]{arXiv:#2}

\bibitem[Acher et~al\mbox{.}(2019)]%
        {acher}
\bibfield{author}{\bibinfo{person}{Mathieu Acher}, \bibinfo{person}{Hugo
  Martin}, \bibinfo{person}{Juliana~Alves Pereira}, \bibinfo{person}{Arnaud
  Blouin}, \bibinfo{person}{Jean-Marc J{\'e}z{\'e}quel},
  \bibinfo{person}{Djamel~Eddine Khelladi}, \bibinfo{person}{Luc Lesoil}, {and}
  \bibinfo{person}{Olivier Barais}.} \bibinfo{year}{2019}\natexlab{}.
\newblock \bibinfo{booktitle}{\emph{{Learning Very Large Configuration Spaces:
  What Matters for Linux Kernel Sizes}}}.
\newblock \bibinfo{type}{Research Report}. \bibinfo{institution}{{Inria Rennes
  - Bretagne Atlantique}}.
\newblock
\urldef\tempurl%
\url{https://inria.hal.science/hal-02314830}
\showURL{%
\tempurl}


\bibitem[Ahmed et~al\mbox{.}(2023)]%
        {LLM_cloud}
\bibfield{author}{\bibinfo{person}{Toufique Ahmed}, \bibinfo{person}{Supriyo
  Ghosh}, \bibinfo{person}{Chetan Bansal}, \bibinfo{person}{Thomas Zimmermann},
  \bibinfo{person}{Xuchao Zhang}, {and} \bibinfo{person}{Saravan Rajmohan}.}
  \bibinfo{year}{2023}\natexlab{}.
\newblock \showarticletitle{Recommending Root-Cause and Mitigation Steps for
  Cloud Incidents Using Large Language Models}. In
  \bibinfo{booktitle}{\emph{Proceedings of the 45th International Conference on
  Software Engineering}} (Melbourne, Victoria, Australia)
  \emph{(\bibinfo{series}{ICSE '23})}. \bibinfo{publisher}{IEEE Press},
  \bibinfo{pages}{1737–1749}.
\newblock
\showISBNx{9781665457019}
\href{https://doi.org/10.1109/ICSE48619.2023.00149}{doi:\nolinkurl{10.1109/ICSE48619.2023.00149}}


\bibitem[Akopytov(2004)]%
        {sysbench}
\bibfield{author}{\bibinfo{person}{Alexey Akopytov}.}
  \bibinfo{year}{2004}\natexlab{}.
\newblock \bibinfo{title}{sysbench: Scriptable database and system performance
  benchmark (version 1.0.20)}.
\newblock
\urldef\tempurl%
\url{https://github.com/akopytov/sysbench}
\showURL{%
\tempurl}


\bibitem[{Apache Software Foundation}(1995)]%
        {Apache}
\bibfield{author}{\bibinfo{person}{{Apache Software Foundation}}.}
  \bibinfo{year}{1995}\natexlab{}.
\newblock \bibinfo{title}{Apache HTTP Server (version 2.4.52)}.
\newblock
\urldef\tempurl%
\url{https://httpd.apache.org/}
\showURL{%
\tempurl}


\bibitem[{Apache Software Foundation}(1997)]%
        {apache_bench}
\bibfield{author}{\bibinfo{person}{{Apache Software Foundation}}.}
  \bibinfo{year}{1997}\natexlab{}.
\newblock \bibinfo{title}{ab - Apache HTTP server benchmarking tool (version
  2.3)}.
\newblock
\urldef\tempurl%
\url{https://httpd.apache.org/docs/2.4/programs/ab.html}
\showURL{%
\tempurl}


\bibitem[Brown et~al\mbox{.}(2020)]%
        {fewshot}
\bibfield{author}{\bibinfo{person}{Tom~B. Brown}, \bibinfo{person}{Benjamin
  Mann}, \bibinfo{person}{Nick Ryder}, \bibinfo{person}{Melanie Subbiah},
  \bibinfo{person}{Jared Kaplan}, \bibinfo{person}{Prafulla Dhariwal},
  \bibinfo{person}{Arvind Neelakantan}, \bibinfo{person}{Pranav Shyam},
  \bibinfo{person}{Girish Sastry}, \bibinfo{person}{Amanda Askell},
  \bibinfo{person}{Sandhini Agarwal}, \bibinfo{person}{Ariel Herbert-Voss},
  \bibinfo{person}{Gretchen Krueger}, \bibinfo{person}{Tom Henighan},
  \bibinfo{person}{Rewon Child}, \bibinfo{person}{Aditya Ramesh},
  \bibinfo{person}{Daniel~M. Ziegler}, \bibinfo{person}{Jeffrey Wu},
  \bibinfo{person}{Clemens Winter}, \bibinfo{person}{Christopher Hesse},
  \bibinfo{person}{Mark Chen}, \bibinfo{person}{Eric Sigler},
  \bibinfo{person}{Mateusz Litwin}, \bibinfo{person}{Scott Gray},
  \bibinfo{person}{Benjamin Chess}, \bibinfo{person}{Jack Clark},
  \bibinfo{person}{Christopher Berner}, \bibinfo{person}{Sam McCandlish},
  \bibinfo{person}{Alec Radford}, \bibinfo{person}{Ilya Sutskever}, {and}
  \bibinfo{person}{Dario Amodei}.} \bibinfo{year}{2020}\natexlab{}.
\newblock \bibinfo{title}{Language Models are Few-Shot Learners}.
\newblock
\showeprint[arxiv]{2005.14165}~[cs.CL]
\urldef\tempurl%
\url{https://arxiv.org/abs/2005.14165}
\showURL{%
\tempurl}


\bibitem[{Byte UnixBench Developers}(1983)]%
        {unixbench}
\bibfield{author}{\bibinfo{person}{{Byte UnixBench Developers}}.}
  \bibinfo{year}{1983}\natexlab{}.
\newblock \bibinfo{title}{UnixBench (version 5.1.3)}.
\newblock
\urldef\tempurl%
\url{https://github.com/kdlucas/byte-unixbench}
\showURL{%
\tempurl}


\bibitem[Chen et~al\mbox{.}(2024)]%
        {autoos}
\bibfield{author}{\bibinfo{person}{Huilai Chen}, \bibinfo{person}{Yuanbo Wen},
  \bibinfo{person}{Limin Cheng}, \bibinfo{person}{Shouxu Kuang},
  \bibinfo{person}{Yumeng Liu}, \bibinfo{person}{Weijia Li},
  \bibinfo{person}{Ling Li}, \bibinfo{person}{Rui Zhang},
  \bibinfo{person}{Xinkai Song}, \bibinfo{person}{Wei Li}, \bibinfo{person}{Qi
  Guo}, {and} \bibinfo{person}{Yunji Chen}.} \bibinfo{year}{2024}\natexlab{}.
\newblock \showarticletitle{AutoOS: Make Your {OS} More Powerful by Exploiting
  Large Language Models}. In \bibinfo{booktitle}{\emph{Forty-first
  International Conference on Machine Learning, {ICML} 2024, Vienna, Austria,
  July 21-27, 2024}}. \bibinfo{publisher}{OpenReview.net}.
\newblock
\urldef\tempurl%
\url{https://openreview.net/forum?id=Rp8R9C0Sth}
\showURL{%
\tempurl}


\bibitem[DeepSeek-AI(2025)]%
        {deepseekai2025deepseekv3technicalreport}
\bibfield{author}{\bibinfo{person}{DeepSeek-AI}.}
  \bibinfo{year}{2025}\natexlab{}.
\newblock \bibinfo{title}{DeepSeek-V3 Technical Report}.
\newblock
\showeprint[arxiv]{2412.19437}~[cs.CL]
\urldef\tempurl%
\url{https://arxiv.org/abs/2412.19437}
\showURL{%
\tempurl}


\bibitem[Evang and Dreibholz(2024)]%
        {NetOptimize1}
\bibfield{author}{\bibinfo{person}{Jan~Marius Evang} {and}
  \bibinfo{person}{Thomas Dreibholz}.} \bibinfo{year}{2024}\natexlab{}.
\newblock \showarticletitle{Optimizing Network Latency: Unveiling the Impact of
  Reflection Server Tuning}. In \bibinfo{booktitle}{\emph{International
  Conference on Advanced Information Networking and Applications}}. Springer,
  \bibinfo{pages}{374--384}.
\newblock


\bibitem[Franz et~al\mbox{.}(2021)]%
        {ConfigFix}
\bibfield{author}{\bibinfo{person}{Patrick Franz}, \bibinfo{person}{Thorsten
  Berger}, \bibinfo{person}{Ibrahim Fayaz}, \bibinfo{person}{Sarah Nadi}, {and}
  \bibinfo{person}{Evgeny Groshev}.} \bibinfo{year}{2021}\natexlab{}.
\newblock \showarticletitle{ConfigFix: Interactive Configuration Conflict
  Resolution for the Linux Kernel}. In \bibinfo{booktitle}{\emph{2021 IEEE/ACM
  43rd International Conference on Software Engineering: Software Engineering
  in Practice (ICSE-SEIP)}}. \bibinfo{pages}{91--100}.
\newblock
\href{https://doi.org/10.1109/ICSE-SEIP52600.2021.00018}{doi:\nolinkurl{10.1109/ICSE-SEIP52600.2021.00018}}


\bibitem[Ghosh et~al\mbox{.}(2025)]%
        {LLM_cve}
\bibfield{author}{\bibinfo{person}{Rikhiya Ghosh}, \bibinfo{person}{Hans-Martin
  von Stockhausen}, \bibinfo{person}{Martin Schmitt},
  \bibinfo{person}{George~Marica Vasile}, \bibinfo{person}{Sanjeev~Kumar Karn},
  {and} \bibinfo{person}{Oladimeji Farri}.} \bibinfo{year}{2025}\natexlab{}.
\newblock \showarticletitle{CVE-LLM: Ontology-Assisted Automatic Vulnerability
  Evaluation Using Large Language Models}. In
  \bibinfo{booktitle}{\emph{Proceedings of the AAAI Conference on Artificial
  Intelligence}}, Vol.~\bibinfo{volume}{39}. \bibinfo{pages}{28757--28765}.
\newblock


\bibitem[Group(1996)]%
        {PostgreSQL}
\bibfield{author}{\bibinfo{person}{PostgreSQL Global~Development Group}.}
  \bibinfo{year}{1996}\natexlab{}.
\newblock \bibinfo{title}{PostgreSQL: The world's most advanced open source
  database (version 14.15)}.
\newblock
\urldef\tempurl%
\url{https://www.postgresql.org/}
\showURL{%
\tempurl}


\bibitem[Guo et~al\mbox{.}(2025)]%
        {lightrag}
\bibfield{author}{\bibinfo{person}{Zirui Guo}, \bibinfo{person}{Lianghao Xia},
  \bibinfo{person}{Yanhua Yu}, \bibinfo{person}{Tu Ao}, {and}
  \bibinfo{person}{Chao Huang}.} \bibinfo{year}{2025}\natexlab{}.
\newblock \bibinfo{title}{LightRAG: Simple and Fast Retrieval-Augmented
  Generation}.
\newblock
\showeprint[arxiv]{2410.05779}~[cs.IR]
\urldef\tempurl%
\url{https://arxiv.org/abs/2410.05779}
\showURL{%
\tempurl}


\bibitem[Ha and Zhang(2019)]%
        {DeepPerf}
\bibfield{author}{\bibinfo{person}{Huong Ha} {and} \bibinfo{person}{Hongyu
  Zhang}.} \bibinfo{year}{2019}\natexlab{}.
\newblock \showarticletitle{DeepPerf: Performance Prediction for Configurable
  Software with Deep Sparse Neural Network}. In \bibinfo{booktitle}{\emph{2019
  IEEE/ACM 41st International Conference on Software Engineering (ICSE)}}.
\newblock
\href{https://doi.org/10.1109/icse.2019.00113}{doi:\nolinkurl{10.1109/icse.2019.00113}}


\bibitem[Hao et~al\mbox{.}(2021)]%
        {JOIE}
\bibfield{author}{\bibinfo{person}{Junheng Hao}, \bibinfo{person}{Muhao Chen},
  \bibinfo{person}{Wenchao Yu}, \bibinfo{person}{Yizhou Sun}, {and}
  \bibinfo{person}{Wei Wang}.} \bibinfo{year}{2021}\natexlab{}.
\newblock \bibinfo{title}{Universal Representation Learning of Knowledge Bases
  by Jointly Embedding Instances and Ontological Concepts}.
\newblock
\showeprint[arxiv]{2103.08115}~[cs.AI]
\urldef\tempurl%
\url{https://arxiv.org/abs/2103.08115}
\showURL{%
\tempurl}


\bibitem[Jiang et~al\mbox{.}(2024)]%
        {CodeSFT}
\bibfield{author}{\bibinfo{person}{Nan Jiang}, \bibinfo{person}{Xiaopeng Li},
  \bibinfo{person}{Shiqi Wang}, \bibinfo{person}{Qiang Zhou},
  \bibinfo{person}{Soneya Hossain}, \bibinfo{person}{Baishakhi Ray},
  \bibinfo{person}{Varun Kumar}, \bibinfo{person}{Xiaofei Ma}, {and}
  \bibinfo{person}{Anoop Deoras}.} \bibinfo{year}{2024}\natexlab{}.
\newblock \showarticletitle{LeDex: Training LLMs to Better Self-Debug and
  Explain Code}.
\newblock \bibinfo{journal}{\emph{Advances in Neural Information Processing
  Systems}}  \bibinfo{volume}{37} (\bibinfo{year}{2024}),
  \bibinfo{pages}{35517--35543}.
\newblock


\bibitem[Jung et~al\mbox{.}(2021)]%
        {Wayfinder}
\bibfield{author}{\bibinfo{person}{Alexander Jung}, \bibinfo{person}{Hugo
  Lefeuvre}, \bibinfo{person}{Charalampos Rotsos}, \bibinfo{person}{Pierre
  Olivier}, \bibinfo{person}{Daniel O\~{n}oro Rubio}, \bibinfo{person}{Felipe
  Huici}, {and} \bibinfo{person}{Mathias Niepert}.}
  \bibinfo{year}{2021}\natexlab{}.
\newblock \showarticletitle{Wayfinder: towards automatically deriving optimal
  OS configurations}. In \bibinfo{booktitle}{\emph{Proceedings of the 12th ACM
  SIGOPS Asia-Pacific Workshop on Systems}} (Hong Kong, China)
  \emph{(\bibinfo{series}{APSys '21})}. \bibinfo{publisher}{Association for
  Computing Machinery}, \bibinfo{address}{New York, NY, USA},
  \bibinfo{pages}{115–122}.
\newblock
\showISBNx{9781450386982}
\href{https://doi.org/10.1145/3476886.3477506}{doi:\nolinkurl{10.1145/3476886.3477506}}


\bibitem[Kroah-Hartman(2019)]%
        {Kernel}
\bibfield{author}{\bibinfo{person}{Greg Kroah-Hartman}.}
  \bibinfo{year}{2019}\natexlab{}.
\newblock \bibinfo{title}{Linux Kernel Release Model}.
\newblock
\urldef\tempurl%
\url{http://kroah.com/log/blog/2018/02/05/linux-kernel-release-model/}
\showURL{%
\tempurl}
\newblock
\shownote{Accessed: 2023-10-05}.


\bibitem[Kroth et~al\mbox{.}(2024)]%
        {MLOS}
\bibfield{author}{\bibinfo{person}{Brian Kroth}, \bibinfo{person}{Sergiy
  Matusevych}, \bibinfo{person}{Rana Alotaibi}, \bibinfo{person}{Yiwen Zhu},
  \bibinfo{person}{Anja Gruenheid}, {and} \bibinfo{person}{Yuanyuan Tian}.}
  \bibinfo{year}{2024}\natexlab{}.
\newblock \showarticletitle{MLOS in Action: Bridging the Gap Between
  Experimentation and Auto-Tuning in the Cloud}.
\newblock \bibinfo{journal}{\emph{Proceedings of the VLDB Endowment}}
  \bibinfo{volume}{17}, \bibinfo{number}{12} (\bibinfo{year}{2024}),
  \bibinfo{pages}{4269--4272}.
\newblock


\bibitem[Kuo et~al\mbox{.}(2022)]%
        {Debloating}
\bibfield{author}{\bibinfo{person}{Hsuan-Chi Kuo}, \bibinfo{person}{Jianyan
  Chen}, \bibinfo{person}{Sibin Mohan}, {and} \bibinfo{person}{Tianyin Xu}.}
  \bibinfo{year}{2022}\natexlab{}.
\newblock \showarticletitle{Set the configuration for the heart of the OS: on
  the practicality of operating system kernel debloating}.
\newblock \bibinfo{journal}{\emph{Commun. ACM}} \bibinfo{volume}{65},
  \bibinfo{number}{5} (\bibinfo{date}{April} \bibinfo{year}{2022}),
  \bibinfo{pages}{101–109}.
\newblock
\showISSN{0001-0782}
\href{https://doi.org/10.1145/3524301}{doi:\nolinkurl{10.1145/3524301}}


\bibitem[Li et~al\mbox{.}(2022)]%
        {science}
\bibfield{author}{\bibinfo{person}{Yujia Li}, \bibinfo{person}{David Choi},
  \bibinfo{person}{Junyoung Chung}, \bibinfo{person}{Nate Kushman},
  \bibinfo{person}{Julian Schrittwieser}, \bibinfo{person}{Rémi Leblond},
  \bibinfo{person}{Tom Eccles}, \bibinfo{person}{James Keeling},
  \bibinfo{person}{Felix Gimeno}, \bibinfo{person}{Agustin Dal~Lago},
  \bibinfo{person}{Thomas Hubert}, \bibinfo{person}{Peter Choy},
  \bibinfo{person}{Cyprien de Masson~d’Autume}, \bibinfo{person}{Igor
  Babuschkin}, \bibinfo{person}{Xinyun Chen}, \bibinfo{person}{Po-Sen Huang},
  \bibinfo{person}{Johannes Welbl}, \bibinfo{person}{Sven Gowal},
  \bibinfo{person}{Alexey Cherepanov}, \bibinfo{person}{James Molloy},
  \bibinfo{person}{Daniel~J. Mankowitz}, \bibinfo{person}{Esme
  Sutherland~Robson}, \bibinfo{person}{Pushmeet Kohli}, \bibinfo{person}{Nando
  de Freitas}, \bibinfo{person}{Koray Kavukcuoglu}, {and}
  \bibinfo{person}{Oriol Vinyals}.} \bibinfo{year}{2022}\natexlab{}.
\newblock \showarticletitle{Competition-level code generation with AlphaCode}.
\newblock \bibinfo{journal}{\emph{Science}} \bibinfo{volume}{378},
  \bibinfo{number}{6624} (\bibinfo{date}{Dec.} \bibinfo{year}{2022}),
  \bibinfo{pages}{1092–1097}.
\newblock
\showISSN{1095-9203}
\href{https://doi.org/10.1126/science.abq1158}{doi:\nolinkurl{10.1126/science.abq1158}}


\bibitem[Lian et~al\mbox{.}(2024)]%
        {Ciri}
\bibfield{author}{\bibinfo{person}{Xinyu Lian}, \bibinfo{person}{Yinfang Chen},
  \bibinfo{person}{Runxiang Cheng}, \bibinfo{person}{Jie Huang},
  \bibinfo{person}{Parth Thakkar}, \bibinfo{person}{Minjia Zhang}, {and}
  \bibinfo{person}{Tianyin Xu}.} \bibinfo{year}{2024}\natexlab{}.
\newblock \bibinfo{title}{Configuration Validation with Large Language Models}.
\newblock
\showeprint[arxiv]{2310.09690}~[cs.SE]
\urldef\tempurl%
\url{https://arxiv.org/abs/2310.09690}
\showURL{%
\tempurl}


\bibitem[Liu et~al\mbox{.}(2024)]%
        {CodePEFT2}
\bibfield{author}{\bibinfo{person}{Bingchang Liu}, \bibinfo{person}{Chaoyu
  Chen}, \bibinfo{person}{Zi Gong}, \bibinfo{person}{Cong Liao},
  \bibinfo{person}{Huan Wang}, \bibinfo{person}{Zhichao Lei},
  \bibinfo{person}{Ming Liang}, \bibinfo{person}{Dajun Chen},
  \bibinfo{person}{Min Shen}, \bibinfo{person}{Hailian Zhou}, {et~al\mbox{.}}}
  \bibinfo{year}{2024}\natexlab{}.
\newblock \showarticletitle{Mftcoder: Boosting code llms with multitask
  fine-tuning}. In \bibinfo{booktitle}{\emph{Proceedings of the 30th ACM SIGKDD
  Conference on Knowledge Discovery and Data Mining}}.
  \bibinfo{pages}{5430--5441}.
\newblock


\bibitem[Luo et~al\mbox{.}(2023)]%
        {DHGE}
\bibfield{author}{\bibinfo{person}{Haoran Luo}, \bibinfo{person}{Haihong E},
  \bibinfo{person}{Ling Tan}, \bibinfo{person}{Gengxian Zhou},
  \bibinfo{person}{Tianyu Yao}, {and} \bibinfo{person}{Kaiyang Wan}.}
  \bibinfo{year}{2023}\natexlab{}.
\newblock \showarticletitle{DHGE: Dual-View Hyper-Relational Knowledge Graph
  Embedding for Link Prediction and Entity Typing}.
\newblock \bibinfo{journal}{\emph{Proceedings of the AAAI Conference on
  Artificial Intelligence}} \bibinfo{volume}{37}, \bibinfo{number}{5}
  (\bibinfo{date}{June} \bibinfo{year}{2023}), \bibinfo{pages}{6467–6474}.
\newblock
\showISSN{2159-5399}
\href{https://doi.org/10.1609/aaai.v37i5.25795}{doi:\nolinkurl{10.1609/aaai.v37i5.25795}}


\bibitem[Luo et~al\mbox{.}(2024a)]%
        {ChatKBQA}
\bibfield{author}{\bibinfo{person}{Haoran Luo}, \bibinfo{person}{Haihong E},
  \bibinfo{person}{Zichen Tang}, \bibinfo{person}{Shiyao Peng},
  \bibinfo{person}{Yikai Guo}, \bibinfo{person}{Wentai Zhang},
  \bibinfo{person}{Chenghao Ma}, \bibinfo{person}{Guanting Dong},
  \bibinfo{person}{Meina Song}, \bibinfo{person}{Wei Lin},
  \bibinfo{person}{Yifan Zhu}, {and} \bibinfo{person}{Anh~Tuan Luu}.}
  \bibinfo{year}{2024}\natexlab{a}.
\newblock \showarticletitle{ChatKBQA: A Generate-then-Retrieve Framework for
  Knowledge Base Question Answering with Fine-tuned Large Language Models}. In
  \bibinfo{booktitle}{\emph{Findings of the Association for Computational
  Linguistics ACL 2024}}. \bibinfo{publisher}{Association for Computational
  Linguistics}, \bibinfo{pages}{2039–2056}.
\newblock
\href{https://doi.org/10.18653/v1/2024.findings-acl.122}{doi:\nolinkurl{10.18653/v1/2024.findings-acl.122}}


\bibitem[Luo et~al\mbox{.}(2024b)]%
        {RoG}
\bibfield{author}{\bibinfo{person}{Linhao Luo}, \bibinfo{person}{Yuan-Fang Li},
  \bibinfo{person}{Gholamreza Haffari}, {and} \bibinfo{person}{Shirui Pan}.}
  \bibinfo{year}{2024}\natexlab{b}.
\newblock \bibinfo{title}{Reasoning on Graphs: Faithful and Interpretable Large
  Language Model Reasoning}.
\newblock
\showeprint[arxiv]{2310.01061}~[cs.CL]
\urldef\tempurl%
\url{https://arxiv.org/abs/2310.01061}
\showURL{%
\tempurl}


\bibitem[Martin et~al\mbox{.}(2021)]%
        {martin2021transfer}
\bibfield{author}{\bibinfo{person}{Hugo Martin}, \bibinfo{person}{Mathieu
  Acher}, \bibinfo{person}{Juliana~Alves Pereira}, \bibinfo{person}{Luc
  Lesoil}, \bibinfo{person}{Jean-Marc J{\'e}z{\'e}quel}, {and}
  \bibinfo{person}{Djamel~Eddine Khelladi}.} \bibinfo{year}{2021}\natexlab{}.
\newblock \showarticletitle{Transfer learning across variants and versions: The
  case of linux kernel size}.
\newblock \bibinfo{journal}{\emph{IEEE Transactions on Software Engineering}}
  \bibinfo{volume}{48}, \bibinfo{number}{11} (\bibinfo{year}{2021}),
  \bibinfo{pages}{4274--4290}.
\newblock


\bibitem[Mortara and Collet(2021)]%
        {dependency}
\bibfield{author}{\bibinfo{person}{Johann Mortara} {and}
  \bibinfo{person}{Philippe Collet}.} \bibinfo{year}{2021}\natexlab{}.
\newblock \showarticletitle{Capturing the diversity of analyses on the Linux
  kernel variability}. In \bibinfo{booktitle}{\emph{Proceedings of the 25th ACM
  International Systems and Software Product Line Conference - Volume A}}
  (Leicester, United Kingdom) \emph{(\bibinfo{series}{SPLC '21})}.
  \bibinfo{publisher}{Association for Computing Machinery},
  \bibinfo{address}{New York, NY, USA}, \bibinfo{pages}{160–171}.
\newblock
\showISBNx{9781450384698}
\href{https://doi.org/10.1145/3461001.3471151}{doi:\nolinkurl{10.1145/3461001.3471151}}


\bibitem[{Nginx, Inc.}(2004)]%
        {Nginx}
\bibfield{author}{\bibinfo{person}{{Nginx, Inc.}}}
  \bibinfo{year}{2004}\natexlab{}.
\newblock \bibinfo{title}{Nginx: A high-performance web server and reverse
  proxy (version 1.18.0)}.
\newblock
\urldef\tempurl%
\url{https://nginx.org/}
\showURL{%
\tempurl}


\bibitem[OpenAI(2024)]%
        {openai2024gpt4technicalreport}
\bibfield{author}{\bibinfo{person}{OpenAI}.} \bibinfo{year}{2024}\natexlab{}.
\newblock \bibinfo{title}{GPT-4 Technical Report}.
\newblock
\showeprint[arxiv]{2303.08774}~[cs.CL]
\urldef\tempurl%
\url{https://arxiv.org/abs/2303.08774}
\showURL{%
\tempurl}


\bibitem[Ren et~al\mbox{.}(2019)]%
        {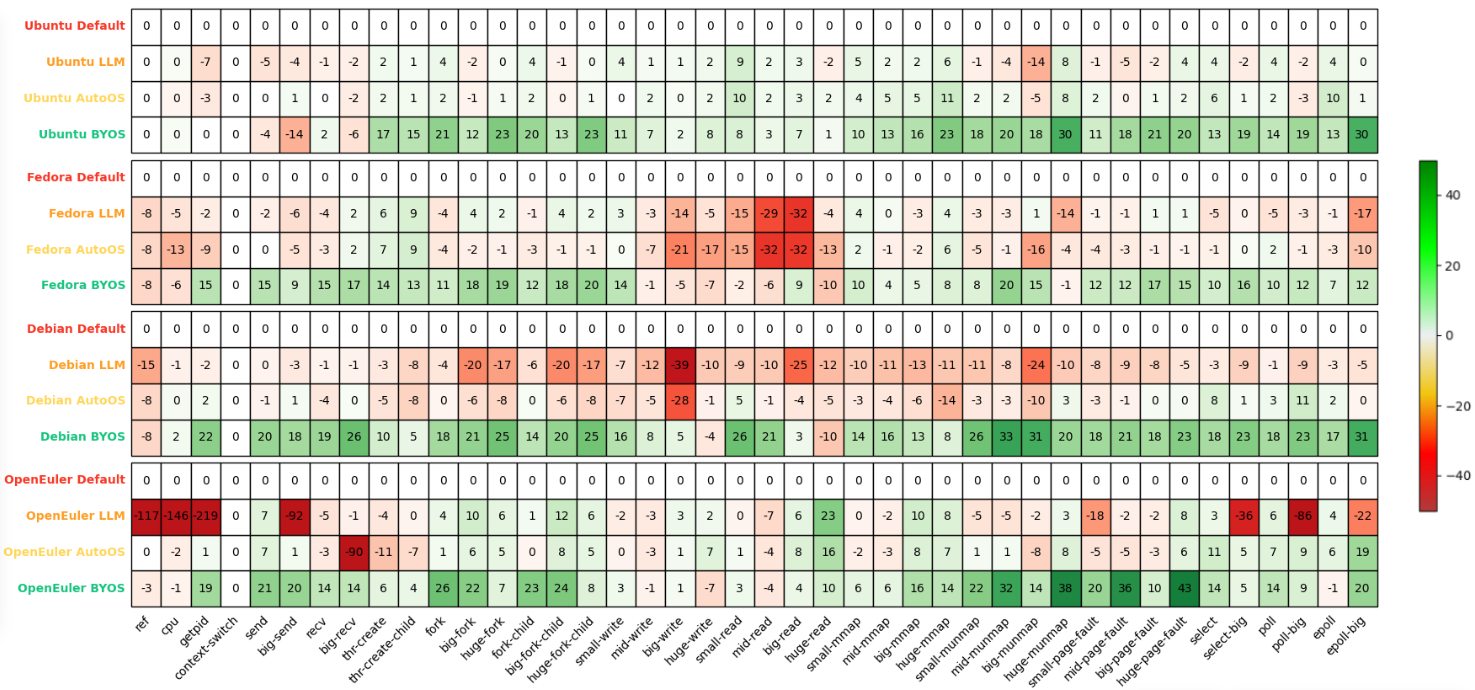}
\bibfield{author}{\bibinfo{person}{Xiang~(Jenny) Ren}, \bibinfo{person}{Kirk
  Rodrigues}, \bibinfo{person}{Luyuan Chen}, \bibinfo{person}{Camilo Vega},
  \bibinfo{person}{Michael Stumm}, {and} \bibinfo{person}{Ding Yuan}.}
  \bibinfo{year}{2019}\natexlab{}.
\newblock \showarticletitle{An analysis of performance evolution of Linux's
  core operations}. In \bibinfo{booktitle}{\emph{Proceedings of the 27th ACM
  Symposium on Operating Systems Principles}} (Huntsville, Ontario, Canada)
  \emph{(\bibinfo{series}{SOSP '19})}. \bibinfo{publisher}{Association for
  Computing Machinery}, \bibinfo{address}{New York, NY, USA},
  \bibinfo{pages}{554–569}.
\newblock
\showISBNx{9781450368735}
\href{https://doi.org/10.1145/3341301.3359640}{doi:\nolinkurl{10.1145/3341301.3359640}}


\bibitem[{Salvatore Sanfilippo}(2009)]%
        {Redis}
\bibfield{author}{\bibinfo{person}{{Salvatore Sanfilippo}}.}
  \bibinfo{year}{2009}\natexlab{}.
\newblock \bibinfo{title}{Redis - The Real-time Data Platform (version
  6.0.16)}.
\newblock
\urldef\tempurl%
\url{https://redis.io/}
\showURL{%
\tempurl}


\bibitem[Sanfilippo(2009)]%
        {redis_benchmarks}
\bibfield{author}{\bibinfo{person}{Salvatore Sanfilippo}.}
  \bibinfo{year}{2009}\natexlab{}.
\newblock \bibinfo{title}{Redis benchmark optimization and management}.
\newblock
\urldef\tempurl%
\url{https://redis.io/docs/latest/operate/oss_and_stack/management/optimization/benchmarks/}
\showURL{%
\tempurl}


\bibitem[Schwarz et~al\mbox{.}(2024)]%
        {NetOptimize2}
\bibfield{author}{\bibinfo{person}{Marcos Schwarz}, \bibinfo{person}{Brian
  Tierney}, \bibinfo{person}{Kiran Vasu}, \bibinfo{person}{Eli Dart},
  \bibinfo{person}{Christian~Esteve Rothenberg}, \bibinfo{person}{Jeronimo
  Bezerra}, {and} \bibinfo{person}{Italo Valcy}.}
  \bibinfo{year}{2024}\natexlab{}.
\newblock \showarticletitle{Recent Linux Improvements that Impact TCP
  Throughput: Insights from R\&E Networks}. In
  \bibinfo{booktitle}{\emph{SC24-W: Workshops of the International Conference
  for High Performance Computing, Networking, Storage and Analysis}}. IEEE,
  \bibinfo{pages}{775--784}.
\newblock


\bibitem[Sun et~al\mbox{.}(2024)]%
        {ToG}
\bibfield{author}{\bibinfo{person}{Jiashuo Sun}, \bibinfo{person}{Chengjin Xu},
  \bibinfo{person}{Lumingyuan Tang}, \bibinfo{person}{Saizhuo Wang},
  \bibinfo{person}{Chen Lin}, \bibinfo{person}{Yeyun Gong},
  \bibinfo{person}{Lionel~M. Ni}, \bibinfo{person}{Heung-Yeung Shum}, {and}
  \bibinfo{person}{Jian Guo}.} \bibinfo{year}{2024}\natexlab{}.
\newblock \bibinfo{title}{Think-on-Graph: Deep and Responsible Reasoning of
  Large Language Model on Knowledge Graph}.
\newblock
\showeprint[arxiv]{2307.07697}~[cs.CL]
\urldef\tempurl%
\url{https://arxiv.org/abs/2307.07697}
\showURL{%
\tempurl}


\bibitem[Tang et~al\mbox{.}(2015)]%
        {paradigm}
\bibfield{author}{\bibinfo{person}{Chunqiang Tang}, \bibinfo{person}{Thawan
  Kooburat}, \bibinfo{person}{Pradeep Venkatachalam}, \bibinfo{person}{Akshay
  Chander}, \bibinfo{person}{Zhe Wen}, \bibinfo{person}{Aravind Narayanan},
  \bibinfo{person}{Patrick Dowell}, {and} \bibinfo{person}{Robert Karl}.}
  \bibinfo{year}{2015}\natexlab{}.
\newblock \showarticletitle{Holistic configuration management at Facebook}. In
  \bibinfo{booktitle}{\emph{Proceedings of the 25th Symposium on Operating
  Systems Principles}} (Monterey, California) \emph{(\bibinfo{series}{SOSP
  '15})}. \bibinfo{publisher}{Association for Computing Machinery},
  \bibinfo{address}{New York, NY, USA}, \bibinfo{pages}{328–343}.
\newblock
\showISBNx{9781450338349}
\href{https://doi.org/10.1145/2815400.2815401}{doi:\nolinkurl{10.1145/2815400.2815401}}


\bibitem[{The Linux Foundation}(2023)]%
        {Kconfig}
\bibfield{author}{\bibinfo{person}{{The Linux Foundation}}.}
  \bibinfo{year}{2023}\natexlab{}.
\newblock \bibinfo{title}{Kconfig Language Documentation}.
\newblock
\urldef\tempurl%
\url{https://docs.kernel.org/kbuild/kconfig-language.html}
\showURL{%
\tempurl}


\bibitem[Wang et~al\mbox{.}(2023)]%
        {reasoning}
\bibfield{author}{\bibinfo{person}{Keheng Wang}, \bibinfo{person}{Feiyu Duan},
  \bibinfo{person}{Sirui Wang}, \bibinfo{person}{Peiguang Li},
  \bibinfo{person}{Yunsen Xian}, \bibinfo{person}{Chuantao Yin},
  \bibinfo{person}{Wenge Rong}, {and} \bibinfo{person}{Zhang Xiong}.}
  \bibinfo{year}{2023}\natexlab{}.
\newblock \bibinfo{title}{Knowledge-Driven CoT: Exploring Faithful Reasoning in
  LLMs for Knowledge-intensive Question Answering}.
\newblock
\showeprint[arxiv]{2308.13259}~[cs.CL]
\urldef\tempurl%
\url{https://arxiv.org/abs/2308.13259}
\showURL{%
\tempurl}


\bibitem[Weyssow et~al\mbox{.}(2023)]%
        {CodePEFT}
\bibfield{author}{\bibinfo{person}{Martin Weyssow}, \bibinfo{person}{Xin Zhou},
  \bibinfo{person}{Kisub Kim}, \bibinfo{person}{David Lo}, {and}
  \bibinfo{person}{Houari Sahraoui}.} \bibinfo{year}{2023}\natexlab{}.
\newblock \showarticletitle{Exploring parameter-efficient fine-tuning
  techniques for code generation with large language models}.
\newblock \bibinfo{journal}{\emph{ACM Transactions on Software Engineering and
  Methodology}} (\bibinfo{year}{2023}).
\newblock


\bibitem[Xia et~al\mbox{.}(2024)]%
        {Agentless}
\bibfield{author}{\bibinfo{person}{Chunqiu~Steven Xia}, \bibinfo{person}{Yinlin
  Deng}, \bibinfo{person}{Soren Dunn}, {and} \bibinfo{person}{Lingming Zhang}.}
  \bibinfo{year}{2024}\natexlab{}.
\newblock \bibinfo{title}{Agentless: Demystifying LLM-based Software
  Engineering Agents}.
\newblock
\showeprint[arxiv]{2407.01489}~[cs.SE]
\urldef\tempurl%
\url{https://arxiv.org/abs/2407.01489}
\showURL{%
\tempurl}


\bibitem[Zhou et~al\mbox{.}(2021)]%
        {KACC}
\bibfield{author}{\bibinfo{person}{Jie Zhou}, \bibinfo{person}{Shengding Hu},
  \bibinfo{person}{Xin Lv}, \bibinfo{person}{Cheng Yang},
  \bibinfo{person}{Zhiyuan Liu}, \bibinfo{person}{Wei Xu}, \bibinfo{person}{Jie
  Jiang}, \bibinfo{person}{Juanzi Li}, {and} \bibinfo{person}{Maosong Sun}.}
  \bibinfo{year}{2021}\natexlab{}.
\newblock \showarticletitle{{KACC}: A Multi-task Benchmark for Knowledge
  Abstraction, Concretization and Completion}. In
  \bibinfo{booktitle}{\emph{Findings of the Association for Computational
  Linguistics: ACL-IJCNLP 2021}}, \bibfield{editor}{\bibinfo{person}{Chengqing
  Zong}, \bibinfo{person}{Fei Xia}, \bibinfo{person}{Wenjie Li}, {and}
  \bibinfo{person}{Roberto Navigli}} (Eds.). \bibinfo{publisher}{Association
  for Computational Linguistics}, \bibinfo{address}{Online},
  \bibinfo{pages}{1751--1763}.
\newblock
\href{https://doi.org/10.18653/v1/2021.findings-acl.153}{doi:\nolinkurl{10.18653/v1/2021.findings-acl.153}}


\end{thebibliography}

\appendix

\section{Benchmark Details}
We evaluate kernel configurations generated by different methods using five representative benchmarks that span computation, memory, storage, networking, and data-intensive workloads.

\textbf{UnixBench} \cite{unixbench} is a general-purpose benchmarking suite for Unix-like systems that measures CPU, memory, and file I/O performance, serving as an overall indicator of system efficiency.

\textbf{LEBench} \cite{LEBench} is a microbenchmark suite that isolates and measures 13 kernel primitives, providing fine-grained insights into how configuration changes affect core OS mechanisms such as scheduling, system calls, and memory management.

\textbf{RedisBench} \cite{redis_benchmarks} simulates concurrent client requests to evaluate the throughput and latency of a Redis server, capturing the characteristics of in-memory key-value workloads sensitive to memory and networking configurations.

\textbf{ApacheBench} \cite{apache_bench} benchmarks HTTP web servers by generating concurrent requests, measuring throughput, latency distribution, and scalability under web-serving workloads.

\textbf{Sysbench} \cite{sysbench} is a modular benchmarking framework; we primarily use its CPU, memory, file I/O, and database modules to assess the impact of kernel tuning on data-intensive applications.

Overall, this benchmark suite enables us to systematically evaluate both general-purpose and workload-specific performance, ensuring that our results are robust, representative, and relevant to real-world deployment scenarios.

\section{BYOS Algorithmic Overview}
\label{sec:alg}

\begin{algorithm}[htbp]
   \caption{Knowledge-driven Configuration Generation in BYOS}
   \label{alg:example}
   \begin{algorithmic}[1]
   \STATE \textbf{Input:} Candidate configuration options $K_q$, OD-KG $\mathcal{G}$, aligned concepts $\mathcal{E}_C^q$
   \STATE \textbf{Output:} Valid kernel configuration $K$
   \STATE \textbf{Step 1: Heuristic Inference for Option Value Assignment.}
   \STATE Initialize $K \gets \emptyset$
   \REPEAT
       \STATE Identify candidate configuration subset $K_t \subseteq K_q$
       \STATE $K_q \gets K_q \setminus K_t$
       
       \FOR{each configuration option $o_t \in K_t$}
           \STATE $x_t \gets \texttt{LLM\_Infer}(o_t \mid \mathcal{E}_C^q, \mathcal{G}, K_t)$
           \IF{$\texttt{Valid}(K_t \cup \{(o_t, x_t)\}) = \texttt{False}$}
               \STATE Prune current assignment.
           \ELSE
               \STATE Add $(o_t, x_t)$ to $K_t$: $K_t = K_t \cup \{(o_t, x_t)\}$
           \ENDIF
       \ENDFOR
   \UNTIL{$K_q = \emptyset$}

   \STATE \textbf{Step 2: Performance-aware Final Configuration Generation.}
   \FOR{each $(o_t, x_t) \in K$}
       \STATE $x_t^* \gets \arg\max\limits_{x \in \mathcal{D}_{o_t}} P(K \cup \{(o_t, x)\}, q)$
       \IF{$\texttt{IsValid}(K \cup \{(o_t, x_t^*)\})$}
           \STATE $K \gets (K \setminus \{(o_t, x_t)\}) \cup \{(o_t, x_t^*)\}$
       \ENDIF
   \ENDFOR
   \STATE \textbf{return} $K$
   \end{algorithmic} 
\end{algorithm}

\paragraph{Algorithm Overview.}
Algorithm~\ref{alg:example} formalizes the core procedure of \textbf{knowledge-driven configuration generation} in BYOS. Given a tuning objective $q$, the aligned concept set $\mathcal{E}_C^q$ (cf. Eq.~(4)), and a candidate option set $K_q$ extracted from the OD-KG $\mathcal{G}$ (cf. Eq.~(6)), the algorithm constructs a valid kernel configuration $K$ that maximizes the performance objective $P(K,q)$ while respecting all constraints in the configuration space $S=(O,E,C)$ (Definition~1). The procedure comprises two stages: (i) \textbf{knowledge-guided value inference} for structural correctness and semantic alignment, and (ii) \textbf{performance-aware refinement} to further optimize $P(K,q)$.

\paragraph{Step 1: Heuristic Inference for Option Value Assignment.}
Starting from the candidate set $K_q$, BYOS iteratively selects a subset $K_t \subseteq K_q$ and assigns values to each option $o_t \in K_t$ via an LLM-based inference function:
\[
    x_t = \texttt{LLM\_Infer}(o_t \mid \mathcal{E}_C^q, \mathcal{G}, K_t),
\]
where $\mathcal{E}_C^q$ provides semantic guidance, $\mathcal{G}$ encodes domain knowledge, and $K_t$ serves as the current partial configuration.

Each inferred assignment $(o_t,x_t)$ is immediately validated in the configuration space $S=(O,E,C)$ by checking: (i) $x_t \in \mathcal{D}_{o_t}$; (ii) $\texttt{Dependencies}(K_t \cup \{(o_t,x_t)\}, E)$; and (iii) $\texttt{Constraints}(K_t \cup \{(o_t,x_t)\}, C)$. Invalid assignments are pruned, while valid ones are incorporated into $K_t$. This process continues until all options in $K_q$ are processed, ensuring valid intermediate configurations and reducing infeasible LLM outputs.

\paragraph{Step 2: Performance-aware Final Configuration Generation.}
Given a valid configuration $K$ from Step~1, BYOS performs optional local refinement. For each $(o_t,x_t)\in K$, it searches the admissible domain $\mathcal{D}_{o_t}$:
\[
    x_t^* = \arg\max_{x \in \mathcal{D}_{o_t}} 
    \left\{ P(K \cup \{(o_t,x)\}, q) \mid \texttt{Valid}(K \cup \{(o_t,x)\}) \right\}.
\]
If the updated configuration remains valid, $x_t$ is replaced by $x_t^*$; otherwise, the original value is kept.

\paragraph{Outcome.}
The algorithm returns a configuration 
\[
K_T = \bigcup_{t=0}^{T-1} \{(o_t, x_t^*)\},
\]
which is valid in $S=(O,E,C)$, semantically aligned with $\mathcal{E}_C^q$ via $\mathcal{G}$, and explicitly optimized for the objective $P(K,q)$—establishing a structured pipeline from high-level intent to deployable kernel configuration.

\section{OD-KG Construction Details}
\label{sec:construct}

\begin{figure*}[htbp]
\begin{center}
\centerline{\includegraphics[width=\textwidth]{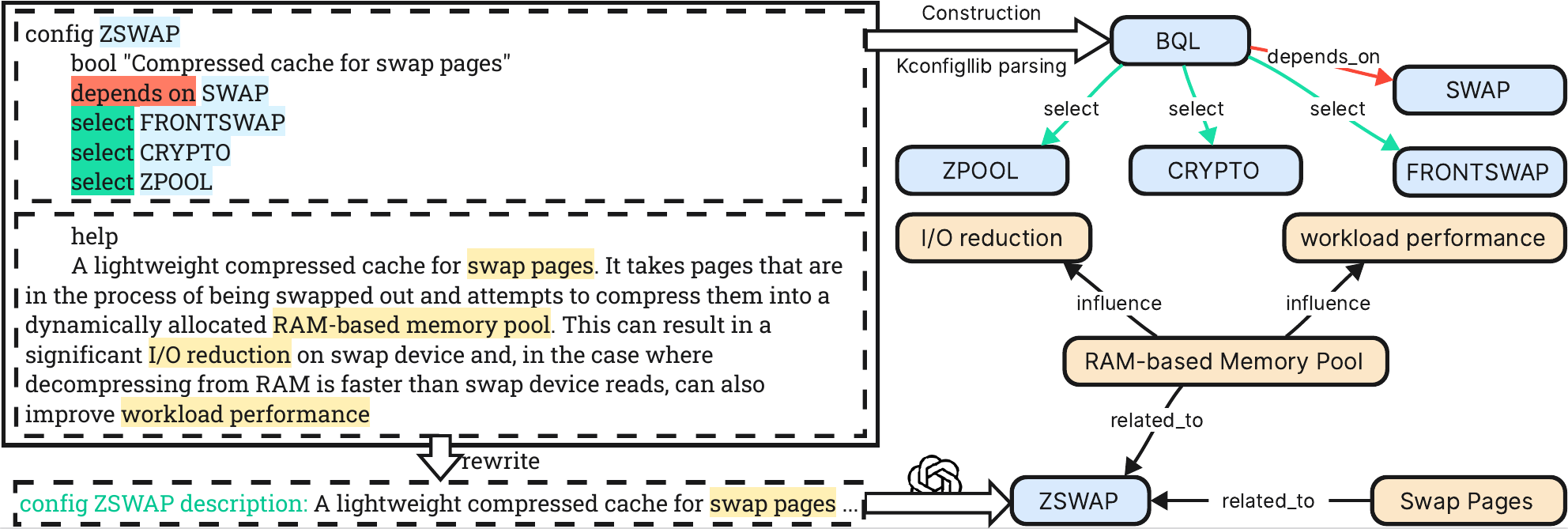}}
\caption{Entity \& Relation Extraction Process}
\label{fig:entity_extraction}
\end{center}
\end{figure*}

\begin{figure*}[!htbp]
\begin{center}
\centerline{\includegraphics[width=\textwidth]{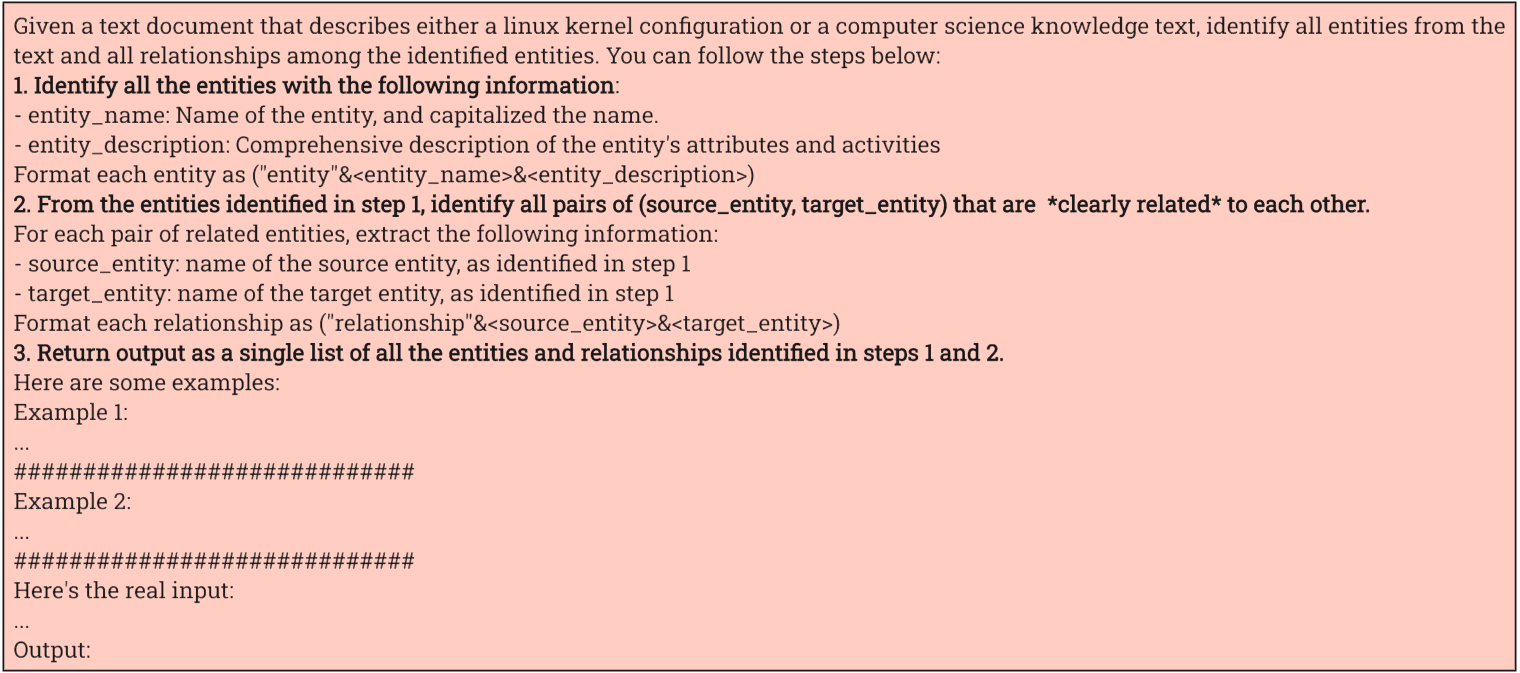}}
\caption{Entity \& Relation Extraction Prompt.}
\label{entity_extraction_prompt}
\end{center}
\end{figure*}

To systematically construct the OS-oriented Dual-layer Knowledge Graph (OD-KG), BYOS extracts entities and relations from both \textbf{structured Kconfig specifications} and \textbf{unstructured textual descriptions}. The overall process is illustrated in Figure~\ref{fig:entity_extraction}.

\paragraph{Structured Kconfig Parsing.}
For structured Kconfig data, we employ \texttt{Kconfiglib} to deterministically parse the official Linux Kconfig specification and construct the \emph{instance layer} of OD-KG. Specifically, each kernel configuration option is mapped to an entity $e \in E_I$, and syntactic dependency expressions in Kconfig are converted into typed relations $r \in R_I$. We cover the four primary Kconfig-defined relations---\texttt{depends\_on}, \texttt{select}, \texttt{imply}, and \texttt{has\_child}---ensuring that structural constraints are faithfully preserved in the graph. 

For example, as shown in Figure~\ref{fig:entity_extraction}, \texttt{CONFIG\_ZSWAP} is represented as an instance-layer entity \textsf{ZSWAP}, with extracted relations such as (\textsf{ZSWAP}, \texttt{depends\_on}, \textsf{SWAP}) and (\textsf{ZSWAP}, \texttt{select}, \textsf{FRONTSWAP}). This deterministic parsing provides a sound and reproducible foundation for subsequent reasoning, reducing the risk of inconsistent or hallucinated structural dependencies.

\paragraph{Textual Entity Detection and Relation Identification.}
Each configuration option is accompanied by \emph{help text} that semantically describes its functionality. To incorporate this unstructured information, we normalize each description into a canonical form:
\[
\text{``Config \textit{<OPTION>} description: \textit{<text>}"}
\]
This normalization ensures consistency across different kernel versions and documentation styles.

We then prompt a LLM to perform joint \textbf{entity detection} and \textbf{relation identification} over these descriptions, extracting concept-layer entities (e.g., \textsf{RAM-based Memory Pool}, \textsf{I/O Reduction}, \textsf{Workload Performance}) and their semantic relations (e.g., \texttt{influence}). These outputs populate the \emph{concept layer} $G_C$ of OD-KG. The prompt template used for this step is shown in Figure~\ref{entity_extraction_prompt}.

\paragraph{Cross-layer Alignment.}
Finally, BYOS employs LLM-based semantic matching to align instance-layer options with concept-layer entities, forming cross-layer links $L = \{(e_I, \texttt{related\_to}, e_C)\}$. These links bridge low-level configuration semantics with high-level tuning objectives, enabling end-to-end interpretability. In Figure~\ref{fig:entity_extraction}, \textsf{ZSWAP} is linked to the concept \textsf{Swap Pages}, which is further connected to \textsf{RAM-based Memory Pool}. This concept is associated with \textsf{I/O Reduction} and \textsf{Workload Performance}, enabling principled reasoning from high-level system objectives to low-level configurations. 

Overall, this construction pipeline—deterministic Kconfig parsing followed by LLM-assisted semantic extraction—ensures that the resulting OD-KG is both \textbf{structurally faithful} to the kernel and \textbf{semantically meaningful} for downstream tuning. This hybrid design balances reliability (from rule-based parsing) and expressiveness (from LLM-based understanding), which is critical for robust and interpretable kernel optimization.

\section{Influence of Different Tuning Prompts}
To assess the sensitivity of BYOS to natural-language descriptions of the tuning objective, we evaluate the impact of different prompt formulations on the resulting configurations and their performance. Specifically, we consider five semantically similar but linguistically distinct descriptions of the optimization goal for Redis:

\textbf{\texttt{P1.}} \texttt{I want to improve the performance of Redis.}

\textbf{\texttt{P2.}} \texttt{Fine-tune Redis for better performance.}

\textbf{\texttt{P3.}} \texttt{I would like to enhance the efficiency of Redis.}

\textbf{\texttt{P4.}} \texttt{Boost the performance of Redis.}

\textbf{\texttt{P5.}} \texttt{My goal is to increase Redis performance.}

For each prompt, we generate a corresponding kernel configuration using BYOS (and the ablated variant without OD-KG) and evaluate its performance using ApacheBench. The results are reported in Table~\ref{tab:new_prompt}.

\begin{table}[!htbp]
\centering
\caption{ApacheBench score of different prompts}
\label{tab:new_prompt}
\resizebox{\columnwidth}{!}{%
    \begin{tabular}{ccccccc|c}
    \toprule
    \textbf{Score (ops/sec)$\backslash$Prompt} 
      & \textbf{P1} & \textbf{P2} & \textbf{P3} & \textbf{P4} & \textbf{P5} \\ 
    \midrule
    \textbf{BYOS}     
      & 189377.98 & 189350.24 & 189370.56 & 189355.20 & 189382.10 \\  
    \textbf{w/o OD-KG}      
      & 155827.86 & 155801.54 & 155827.29 & 155815.60 & 155845.11 \\
    \bottomrule
    \end{tabular}
}
\end{table}

BYOS consistently outperforms the variant without OD-KG across all five prompt formulations, with an average improvement of approximately 21.5\% in throughput. This indicates that the performance gains of BYOS are not artifacts of a particular prompt wording but stem from its knowledge-driven reasoning mechanism.

Moreover, the performance of BYOS remains highly stable across different prompt phrasings, with only minor fluctuations (within 0.02\%) among P1--P5. This robustness suggests that BYOS effectively maps semantically equivalent objectives to similar concept sets $\mathcal{E}_C^q$ via the alignment function $\phi(\cdot)$ (Eq.~(4)), thereby reducing sensitivity to linguistic variations. In contrast, the variant without OD-KG exhibits lower and more unstable performance, implying that direct LLM inference without structured knowledge is more vulnerable to prompt ambiguity.

These results validate the design rationale of BYOS: by grounding natural-language objectives in the OD-KG, BYOS mitigates prompt sensitivity, constrains the search space to semantically relevant configurations, and enables more reliable and consistent tuning outcomes across diverse textual formulations of the same target.

\section{Limitations and Future Work}
\paragraph{Operating System Scope.}
Our current implementation and evaluation focus exclusively on Linux. We did not experiment with other mainstream operating systems such as Windows or macOS because they are largely closed-source and, to the best of our knowledge, do not expose a publicly accessible, structured configuration space comparable to Linux Kconfig. Consequently, the generalizability of BYOS beyond Linux remains limited.

In future work, we will engage with relevant communities to explore whether analogous configuration abstractions or tuning interfaces can be made available for research. If direct access remains infeasible, we will investigate alternative LLM-based tuning strategies (e.g., log- or telemetry-guided optimization) that do not require kernel recompilation.

\paragraph{Experimental Constraints.}
A practical limitation arises from the high cost of kernel compilation and testing: integrating a configuration, rebuilding the kernel, and benchmarking typically takes over an hour per run. This constraint limited the number of configurations we could evaluate on a single machine. Although we parallelized experiments across multiple machines while keeping hardware consistent within each group, this still restricted the sample size, which may not fully capture performance variance.

In future work, we will pursue more efficient evaluation pipelines, including incremental builds, virtualization-based testing, and lightweight performance proxies, to enable larger-scale and more statistically robust studies.

\begin{figure*}[!htbp]
    \centering
    \includegraphics[width=\textwidth]{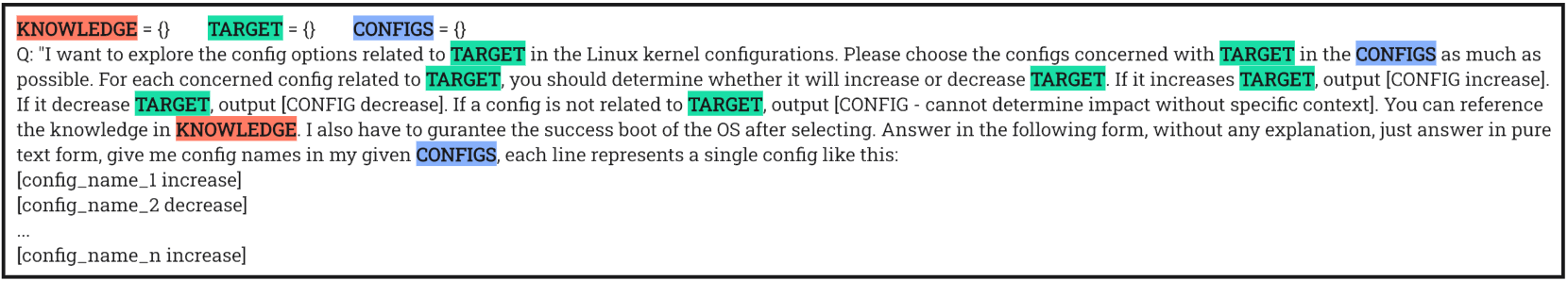}
    \caption{Prompt template for \emph{Bool}-type configuration options assignment.}
    \label{bool_prompt}
\end{figure*}

\begin{figure*}[!htbp]
\begin{center}
\centerline{\includegraphics[width=\textwidth]{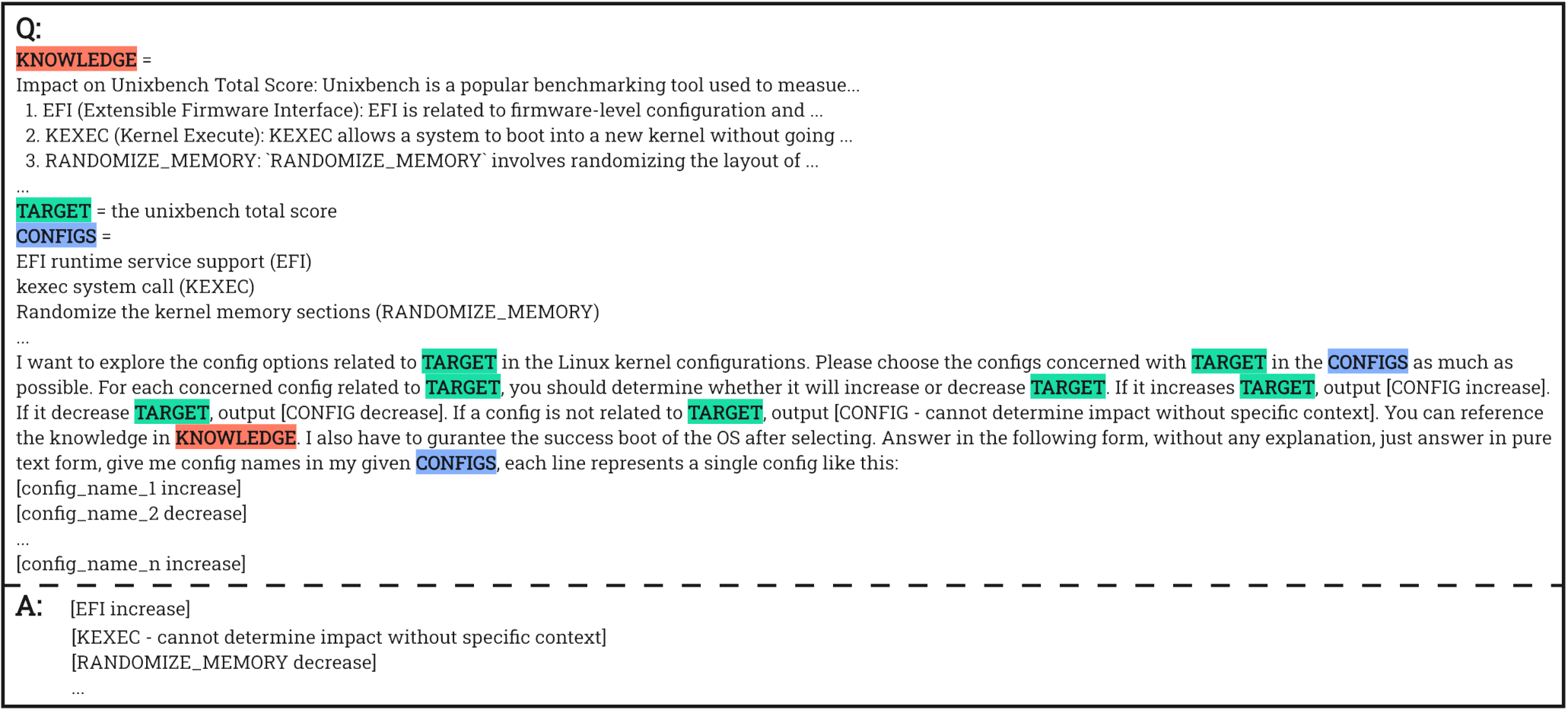}}
\caption{Prompt instance for \emph{Bool}-type configuration assignment.}
\label{bool_query}
\end{center}
\end{figure*}

\section{Type-Specific Prompts in BYOS}
\label{sec:interact}
To effectively explore the Linux kernel configuration space, BYOS interacts with LLMs using type-specific prompts tailored to four Kconfig option types—\textbf{Bool, Choice, Menu, and Value}—each reflecting a distinct decision pattern in kernel tuning. Despite their differences, all prompts share a unified structure comprising: (i) a user-specified tuning objective (\texttt{TARGET}), (ii) external knowledge relevant to \texttt{TARGET} retrieved by LightRAG \cite{lightrag} (\texttt{KNOWLEDGE}), and (iii) a set of candidate configurations (\texttt{CONFIGS} or \texttt{DIRECTORIES}). This design enables consistent, knowledge-grounded reasoning while respecting the semantics of each option type.

\paragraph{Bool Prompts}
Bool options take binary values (\texttt{on/off}). To reduce query cost and stabilize LLM reasoning, we batch up to nine Bool options per query. Rather than directly requesting Boolean assignments, we ask the LLM to infer each option’s \emph{effect} on the target: \texttt{increase}, \texttt{decrease}, or \texttt{cannot determine}. This indirect elicitation mitigates brittle yes/no predictions and encourages explanation-aware reasoning. The prompt template is shown in Figure~\ref{bool_prompt}, and an example query appears in Figure~\ref{bool_query}. 

\begin{figure*}[!h]
    \centering
    \includegraphics[width=\textwidth]{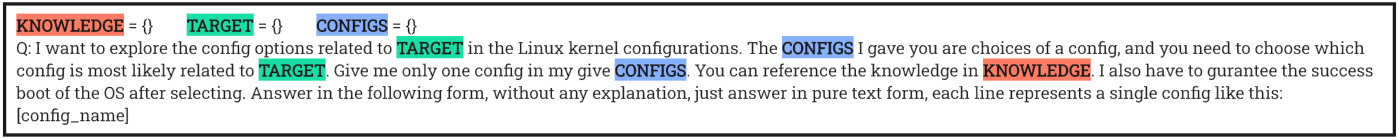}
    \caption{Prompt template for \emph{Choice}-type configuration option selection.}
    \label{choice_prompt}
\end{figure*}

\begin{figure*}[!htbp]
\begin{center}
\centerline{\includegraphics[width=\textwidth]{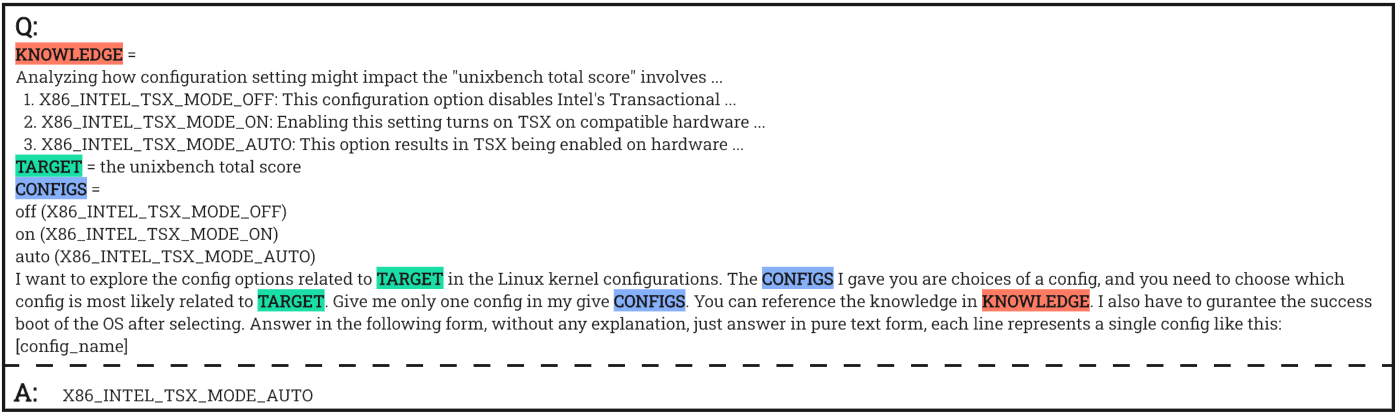}}
\caption{Prompt instance for \emph{Choice}-type configuration selection.}
\label{choice_query}
\end{center}
\end{figure*}

\paragraph{Choice Prompts}
A Choice option contains multiple alternatives, of which exactly one must be selected. Accordingly, the LLM is prompted to choose the most appropriate option given \texttt{KNOWLEDGE} and \texttt{TARGET}. This formulation aligns with the mutual exclusivity semantics of Choice options. The prompt template is shown in Figure~\ref{choice_prompt}, and an example query is in Figure~\ref{choice_query}. 

\begin{figure*}[!h]
    \centering
    \includegraphics[width=\textwidth]{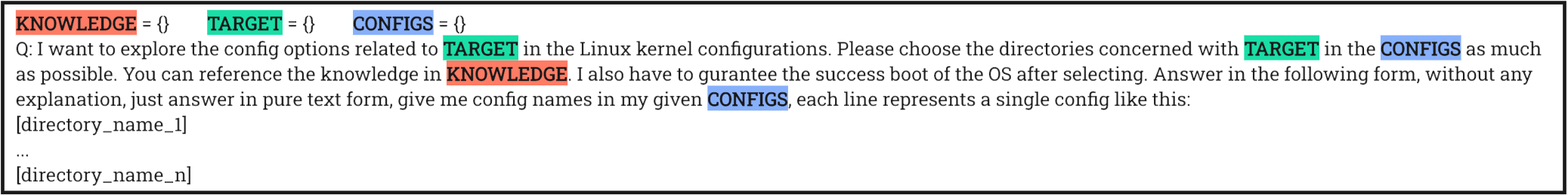}
    \caption{Prompt template for \emph{Menu}-type configuration option selection.}
    \label{menu_prompt}
\end{figure*}

\begin{figure*}[!ht]
\begin{center}
\centerline{\includegraphics[width=\textwidth]{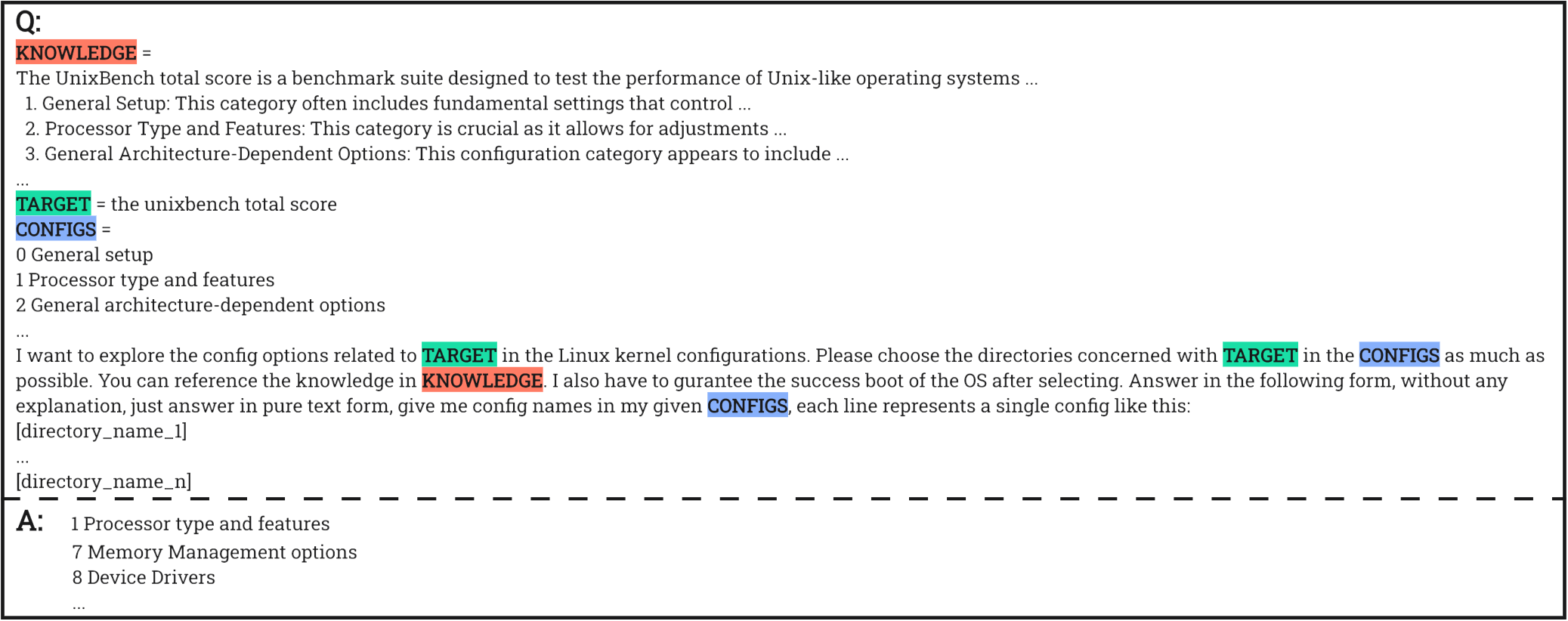}}
\caption{Prompt instance for \emph{Menu}-type configuration selection.}
\label{menu_query}
\end{center}
\end{figure*}

\paragraph{Menu Prompts}
A Menu option encapsulates a hierarchical set of sub-options. Instead of selecting values, we ask the LLM whether the menu contains sub-options potentially relevant to \texttt{TARGET}. If so, the menu is added to the exploration list for further refinement. This enables BYOS to prioritize promising regions of the configuration space while avoiding exhaustive traversal. The prompt template and example are shown in Figures~\ref{menu_prompt} and~\ref{menu_query}, respectively. Here, \texttt{DIRECTORIES} replaces \texttt{CONFIGS} to reflect the hierarchical structure.

\begin{figure*}[!ht]
    \centering
    \includegraphics[width=\textwidth]{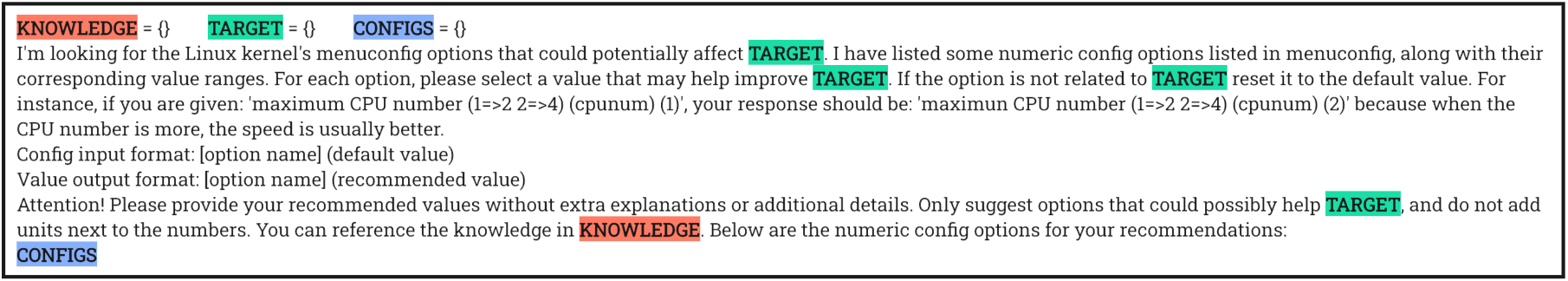}
    \caption{Prompt template for \emph{Value}-type configuration options assignment.}
    \label{value_prompt}
\end{figure*}

\begin{figure*}[!ht]
\begin{center}
\centerline{\includegraphics[width=\textwidth]{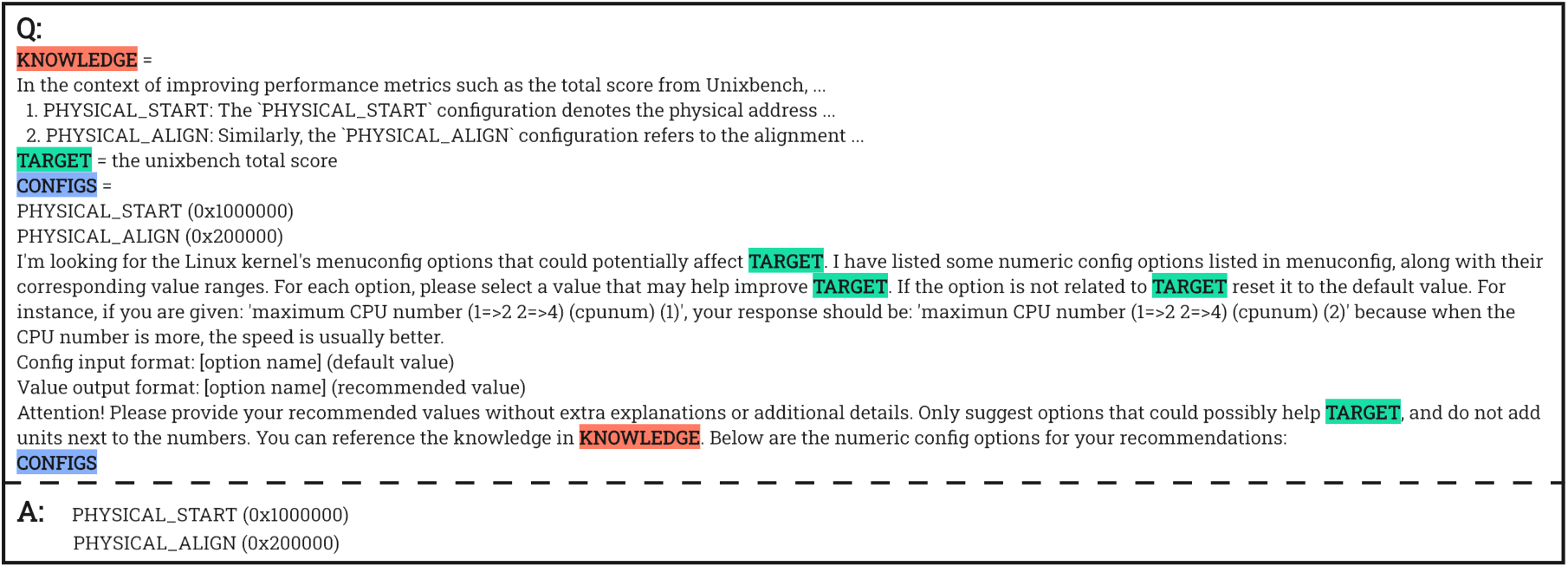}}
\caption{Prompt instance for \emph{Value}-type configuration assignment.}
\label{value_query}
\end{center}
\end{figure*}

\paragraph{Value Prompts}
Value options (e.g., \texttt{int}, \texttt{hex}, or \texttt{string}) require numerical or textual assignments within a valid domain $\mathcal{D}_{o}$. The LLM is prompted to propose candidate values base on \texttt{KNOWLEDGE}, \texttt{TARGET}, and \texttt{CONFIGS}, after which BYOS validates them against kernel constraints. The prompt template is shown in Figure~\ref{value_prompt}, with an example in Figure~\ref{value_query}.

\newpage

\end{document}